\documentclass[12pt,letterpaper]{article}
\usepackage{amsmath,amssymb,pgf,pgfarrows,pgfnodes,float,appendix, hyperref}
\usepackage{graphicx}
\usepackage{subfigure}
\usepackage[margin=0.9in]{geometry}

\newcommand{\nn}{\nonumber}

\newcommand{\be}{\begin{equation}}
\newcommand{\ee}{\end{equation}}
\newcommand{\bea}{\begin{eqnarray}}
\newcommand{\eea}{\end{eqnarray}}

\def\O{{\cal O}}

\title{{\rm\footnotesize \qquad \qquad \qquad \qquad \qquad \ \qquad \qquad \qquad \ \ \ \ \ \                  UTTG-30-13\ TCC-024-13     RUNHETC-2013-17     
SCIPP 13/11}\vskip.5in     Holographic Space-time and Black Holes: Mirages As Alternate Reality}
\author{Tom Banks\\
Department of Physics and SCIPP\\
University of California, Santa Cruz, CA 95064\\
{\it and}\\
Department of Physics and NHETC\\
Rutgers University, Piscataway, NJ 08854\\
E-mail: \href{mailto:banks@scipp.ucsc.edu}{banks@scipp.ucsc.edu}
\\
\\
\centerline{Willy Fischler, Sandipan Kundu and Juan F. Pedraza}\\
Department of Physics and Texas Cosmology Center\\
University of Texas, Austin, TX 78712\\
E-mail: \href{mailto:fischler@physics.utexas.edu}{fischler@physics.utexas.edu}, \href{mailto:sandyk@physics.utexas.edu}{sandyk@physics.utexas.edu}, \\\href{mailto:jpedraza@physics.utexas.edu}{jpedraza@physics.utexas.edu}}

\date{}
\begin{document}
\maketitle

\begin{abstract}
We revisit our investigation of the claim of \cite{amps} that old black holes contain a firewall, {\it i.e.} an in-falling observer encounters highly excited states at a time much shorter than the light crossing time of the Schwarzschild radius. We used the formalism of Holographic Space-time (HST) where there is no dramatic change in particle physics inside the horizon until a time of order the Schwarzschild radius. We correct our description of the interior of the black hole .  HST provides a complete description of the quantum mechanics along any time-like trajectory, even those which fall through the black hole horizon.  The latter are described as alternative factorizations of the description of an external observer, turning the mirage of the interior provided by that observer's membrane paradigm on the stretched horizon, into reality.

\end{abstract}

\section{Introduction}

A recent paper \cite{amps} has made a dramatic new claim about black hole physics, and has led to a burgeoning literature\cite{firewall}.  The classical physics of black holes indicates that an in-falling observer will encounter a singularity in a time of order the light crossing time of the Schwarzschild radius, $R_S$.  While the geometry near the singularity is unstable to small perturbations, the geometry a Schwarzschild radius away is not.  It has therefore been assumed that this conclusion holds for black holes that have evolved from larger holes by Hawking evaporation.

In fact, there is only one extant theory of QG, which discusses the observations of local observers\cite{HST}.  HST describes space-time in terms of {\it an infinite number of independent quantum systems, each of which describes the proper time evolution along a particular time-like trajectory.}  The dynamics along each trajectory defines the quantum notion of a {\it causal diamond} because it couples together only a finite number of degrees of freedom during a finite amount of proper time.  Space-time as a whole is knit together by prescribing overlaps between the causal diamonds along one trajectory and those along others.  These overlaps are tensor factors in the Hilbert spaces describing individual diamonds.  The fundamental dynamical principle of HST is that {\it for each overlap, the density matrix defined by the dynamics along one trajectory, is unitarily equivalent to the density matrix defined by the other trajectory's dynamics.}  

The second major difference between HST and the hypothetical quasi-field theoretical model of QG assumed in most of the firewall literature\cite{firewall} is the distinction between particle degrees of freedom and the rest of the variables of QG.  In \cite{firewall}, DOF that are not well described by QFT are assumed to be ``high energy".  In contrast, as we will describe in the next section, HST assigns $o(N^2 )$ DOF to a causal diamond whose holographic screen has an area of $o(N^2 )$ in Planck units.  At most $o(N^{3/2} )$ of these DOF have a QFT-like character.  The others, which we call {\it horizon DOF}, actually have low energy in the Hamiltonian that describes a geodesic observer in flat space, but are decoupled from the particles as $N\rightarrow\infty$.  

This paper is a major revision of our previous study\cite{fw1} of the firewall paradox in HST.  The description of the interior of a black hole discussed in that paper will be corrected and expanded.  In the new description, no individual trajectory Hamiltonian becomes singular or has Planck scale time dependence.  The singularity of the internal black hole geometry is instead expressed through the overlap conditions, which encode the shrinking transverse cross section and infinite radial expansion of the Schwarzschild geometry. 

What we will propose is, that to leading order in space-time curvature, the external trajectory's description is equivalent, each time particles fall into the black hole, to the model we have proposed for the quantum mechanics of a single geodesic in dS space\cite{holounruh}.  That is, particles propagate for a while and then are absorbed by a thermal bath.  In contrast to dS space, the size of the bath depends on initial conditions in Minkowski space, and can be changed both by new particles falling on the black hole and by Hawking evaporation.  There is {\it nothing} singular in this Hamiltonian.

To describe interior trajectories we simply recognize that the external trajectory's description of the QM of the black hole, allows for alternative tensor factorizations of the black hole Hilbert space, corresponding to the causal structure as measured along such an interior trajectory.  For example, a trajectory crossing the horizon at the same time as some particles, but at a different solid angle on the sphere, will have a causal diamond after horizon crossing, which has no overlap with the particles.  As time goes on, it may encounter the particles before it hits the singularity, or not.  Both alternatives can be identified by describing the relation between DOF included in the causal diamond of the trajectory up to a given proper time $t$, and those describing particles.

In a similar way, one can prescribe Hilbert space overlaps that mirror the causal diamonds of two geodesic trajectories that cross the horizon at the same time but at different angles.  The key point is that, the tensor factors corresponding to the diamonds of these trajectories a short time after horizon crossing have no overlap, but their overlap is complete in a time of order $R_S$.   This is a completely non-singular quantum mechanical description of the shrinking of transverse two spheres down to Planckian radius in the classical black hole geometry.

The infinite radial stretching of the interior geometry is similarly encoded in the fact that trajectories that fall into the black hole with time separation greater than $o(R_S)$, have no overlap.  Since absence of Hilbert space overlap translates into space-like separation in geometry, the geometry must add an infinite amount of space (actually an amount proportional to $R_S^3 M_P^2$ ) in order to incorporate the lack of overlap in the QM description. From the external point of view, there is a simpler explanation. When particles fall into the black hole late, they add new DOF to its QM.  Trajectories that fell into the black hole and equilibrated at an earlier time had no causal contact (inside the horizon) with these new DOF and so there is no overlap between their Hilbert spaces.   The singular stretching of the black hole is the way the dual geometry reflects the underlying QM.

It's clear that the in-falling trajectories and external trajectories provide alternative, but gauge equivalent descriptions, of the process of black hole formation and evaporation.  They are gauge equivalent because, by definition, the full QM Hamiltonian gives the same S-matrix.  The in-falling trajectories are simply alternative tensor factorizations of the same system, which are useful for a finite time if the black hole radius is large.

Many researchers view the description of the interior by the membrane paradigm to be a {\it mirage}.  In our formalism, once the membrane paradigm becomes fully quantum mechanical, the mirage serves as a definition of the interior.  Nowhere in this description does one see a firewall.  Proponents of the firewall have argued that the image of the interior provided by the membrane paradigm cannot prove that there is no firewall.  We agree with this\footnote{Though once Hawking evaporation is taken into account, the simple statement that collisions with firewall particles are causally disconnected from the horizon is no longer true. Nonetheless, in models of
evaporating geometries, these collisions will not make a large classical imprint on the horizon.}. However, the argument for the firewall itself is extremely indirect and uses features of the field theoretic description of black hole entropy, and evaporation, which are not part of HST.  As we will try to make clear, our models ensure that the assumption of smooth evolution for proper times of order $R_S$ along in-falling trajectories, does not conflict with unitarity, approximate locality or other general principles.  We suspect that only one of the large class of models we discuss in this paper gives rise to a Lorentz invariant S-matrix, but they all have primitive versions of particle physics and black hole physics.   The assumption that needs to be dropped to avoid the AMPS paradox, is the validity of the detailed field theoretic description of Hawking evaporation in terms of pair creation.

Throughout this paper, all distances and times will be expressed in Planck units.

\section{HST, Sketchily}

This section is intended to adumbrate the formalism of HST, which has been extensively reviewed in\cite{HST}.  The fundamental elements of the theory are the quantum mechanical descriptions of  single time-like trajectories in space-time.  Each is a (proper) time dependent unitary evolution operator $U(t_{f\ n} , t_{i\ n})$ which incorporates the requirements of causality.  $U$ operates in a Hilbert space associated with the beginning and end of the maximal causal diamond of that trajectory.  At intermediate times, $U$ factorizes into an operator acting on a small tensor factor of this Hilbert space, associated with the smaller causal diamond of this interval of proper time, and one operating on its tensor complement.  The dimensions of these Hilbert spaces, when they are large, determine the areas of the holographic screens of the associated diamonds, according to the BHFSB\cite{BHFSB} formula.

Space-time is knit together by combining an infinite congruence of such time-like trajectories.  For each pair of trajectories, at each time, one must specify an {\it overlap Hilbert space}, ${\cal O}$, whose dimension determines the holoscreen area of the largest causal diamond in the intersection of the diamonds of the individual trajectories.  ${\cal O}$ is a (usually) proper tensor factor of each of the individual diamond Hilbert spaces.   The fundamental dynamical constraint of HST is that the dynamics and initial conditions in each trajectory Hilbert space must be chosen in such a way that the two density matrices in ${\cal O}$ specified by the individual trajectory time evolution operators and initial conditions, are unitarily equivalent for every overlap.  The pattern of overlaps and their dimensions, completely specifies the causal structure and conformal factor of a Lorentzian geometry. The quantum data is an infinite collection of quantum systems, with a causal structure built into the time dependent Hamiltonian of each system.  The causal structure of {\it space-time} is determined by the overlap conditions above.  Together these quantum data determine the causal relations and geometrical sizes of the causal diamonds along a space-filling collection of time-like trajectories in space-time.
One may add trajectories to the collection, corresponding to accelerated frames of reference {\it etc.} as long as all overlap conditions can be satisfied in a consistent manner.

For space-times with four non-compact dimensions the fundamental variables in a causal diamond of area $\sim N^2$ are complex $N \times N + 1$ matrices $\psi_i^A (P)$.  They should be thought of as spinor spherical harmonics on the two-sphere, with $N$ an eigenvalue cutoff on the Dirac operator on the sphere.  Similarly, the label $P$ labels a set of spinor harmonics on an internal manifold, with an independent cutoff on that Dirac operator.  The anti-commutation relations of the variables are represented on a Hilbert space whose dimension is the exponential of the number of $\psi_i^A (P)$, and this requirement implements the BHFSB relation between area and entropy.  For simplicity, in this article, the reader can think of the $\psi_i^A (P)$ variables as having fermionic commutation relations, and that there are only a small number of values of $P$.  This corresponds to a compactification of M-theory on a 7 torus of minimal size, in the sense that the system has the minimal entropy for excitations of the compact space, that is allowed by the rules of quantum gravity.  It is likely that none of our conclusions will be altered for more complicated compactifications.

In the limit $N\rightarrow\infty$ one can show\cite{susy} that the commutation relations contain a subalgebra identical to that of the supersymmetry generators on a Fock space of massless superparticles. In fact, we've recently realized that a better description of the large $N$ limiting Hilbert space comes from interpreting the limiting variables as generators of the super-Bondi-Metzner-Sachs algebra\cite{ags}.  These generators,  describe flows of energy, momentum and other quantum numbers through localized regions on the limiting holographic screen. Individual particles are a singular basis for the space of states spanned by the action of the SBMS algebra, and these singularities can lead to infrared divergent formulae for the transition amplitudes.  The S-matrix maps the representation of the SBMS algebra on past infinity (equivalently the backward light-cone in the space of dual momenta), to that on future infinity (the forward light cone of dual momenta)\footnote{Strominger\cite{andy} has interpreted this intertwining of the two algebras as spontaneous breakdown of the BMS group, viewed as a symmetry group.  Gravitons are the Goldstone bosons. This accords with old parallels pointed out by Weinberg\cite{wein} between low energy theorems for soft photons, gravitons and Goldstone bosons of internal symmetries.}.

The Fock space arises, as is familiar from Matrix Theory\cite{bfss} and other large N matrix models from block diagonal $N\times N$ matrices $\psi_i^A \psi_{ A}^{\dagger\ j} $, with blocks of size $K_i$ that go to infinity at fixed ratio.  When $N$ is finite, this leads to an interesting constraint on the concept of particles.   In order to have an approximate Fock space, we need lots of blocks, but in order to localize the states on the sphere, to give them a particle interpretation, we need large blocks.  The maximal entropy compromise is reached for $K_i \sim \sqrt{N}$, which gives an entropy $N^{3/2}$.  This scaling of particle entropy is well known from black hole physics.  It represents the maximal entropy in particles within a causal diamond, which will ${\it not}$ collapse to form a black hole.  The rest of the DOF do not have a particle interpretation.  We call them {\it horizon DOF}, and they contribute the overwhelming majority of the entropy of the system.  

Particle states are defined by an asymptotic constraint
$$ \psi_i^A | particle \rangle = 0,$$ for $NK + Q$ matrix elements, with $ 1 \ll K,Q \ll N$ . On such states the matrix $$ M^j_k = \psi_{A}^{\dagger\ j} \psi_i^A $$  breaks up into $K_i \times K_i $ blocks, plus a large block of size $N - K \times N - K$.  We have $\sum K_i = K$ and $\sum_{i \neq j} K_i K_j = Q$. The constrained Hilbert space factorizes as a tensor product of Hilbert spaces generated by the the variables in separate blocks, and we will assume that the incoming state on the past boundary of the large causal diamond of size $N$ is a product state.  Since the time evolution is linear, this is not really a restriction, because we can figure out the evolution of any state from the behavior of this basis.

The super-BMS interpretation of the large $N$ limit of the $\psi_i^A$ leads to an interpretation of the constraints\cite{superBMS}.
The physical interpretation of the small blocks is that they represent flows of energy, momentum, spin, and in principle other quantum numbers, through cones\footnote{These are analogous to the cones of Sterman-Weinberg jets.} whose outer boundaries lie on the holographic screen.  The vanishing generators are the variables which live on annuli surrounding each of the cones, and are there to guarantee {\it jet isolation}.  The large block, represents the possibility that the initial and final states, can have arbitrary excitation of DOF that carry no energy through the screen.  The Hamiltonian that we will construct for a causal diamond of finite area $N^2$ has the property that the contribution of these DOF goes to zero as $N \rightarrow\infty$.

Note that the above statement about the tensor product structure is valid for fixed values of $K$ and $Q$.  The full asymptotic Hilbert space is a direct sum of 
spaces for each choice of constraints.   As far as particles go, this is reminiscent of the construction of Fock space\footnote{We will see in a moment that the symmetrization of the product states follows from the form of the Hamiltonian we choose.}  However the size of the large block depends on $K$, and we must respect all of the inequalities in taking the limit.  There will surely be mathematical subtleties to be mastered in order to fully understand this limit.

The Hamiltonian we postulate\cite{hstsmatrix} has the following form 
$$H_{in} (N)  = \sum P_0^i + \frac{1}{N^2} \sum g_k (N) {\rm tr}\ M^k .$$
The couplings $g_k$ approach constants at large $N$ and the order of the polynomial is independent of $N$ and $\geq 7$.  In higher dimensions the order must grow like $N^{d-4}$. This Hamiltonian is defined in each individual constrained sector, and the $P_0^i$ are the generators of the single particle/jet energy operators in each $K_i \times K_i$ block, defined through the SUSY algebra described above.  It should be understood that not all values of the couplings are allowed - they are constrained by the overlap conditions.  This Hamiltonian is the generator of the motion in proper time along a geodesic in Minkowski space, and the particle energies refer to the Lorentz frame in which that geodesic is at rest.   The compatibility constraints of HST take on a familiar form when restricted to all possible Minkowski geodesics.

The asymptotic causal diamonds of these trajectories overlap completely, and coincide with the conformal boundary of Minkowski space.  Thus, the density matrix constraints become constraints on the pure asymptotic states.   The S-matrix, $U(\infty, - \infty)$ should be the same for all trajectories, up to a unitary conjugation.  Since the trajectories are related by Poincare transformations, the asymptotic Hilbert space must carry a unitary representation of the Poincare group, which commutes with the S-matrix.   We expect that the coefficients $g_k (N)$ will be at least highly constrained by this requirement. Indeed, we have briefly discussed the fact that the commutation relations for the spinor variables define the analog of the internal geometry of space-time.  Experience from string theory suggests that these commutation relations will also be highly constrained by the existence of a unitary, Lorentz invariant S-matrix.  We have restricted attention in this article to the simplest consistent spectrum: a maximally supersymmetric compactification to 4D with minimal entropy for the internal DOF.

There is however a subtlety in the above argument, of which we've only recently become aware.  Consider two Minkowski geodesics, at rest w.r.t. each other, and at a space-like separation of $M$ Planck units.   The overlap between their causal diamonds at time $N$, has an entropy deficit of order $MN$.  As $N$ goes to infinity at fixed $M$, the fraction of the entropy in individual diamonds, which is not contained in the overlap, goes to zero.  Thus, for generic states in the Hilbert space, the overlap is maximally entangled with the part of the diamond that is inaccessible to it, and thus all of the quantum information in the system, can be read from the density matrix of the overlap.  This is essentially Page's argument \cite{page}.
Our claim that the S matrices computed along the two trajectories are identical would follow if the states were generic.  However, particle states are, by definition, not generic in HST, so we must be more careful.

Our original argument for identical S-matrices was based, implicitly, on intuition about particle physics.  Consider the same pair of causal diamonds, and a massless particle emitted from one of the trajectories.   At the large time $N$, this particle may well be outside the causal diamond of the second trajectory, but since $N \gg M$, the distance between the trajectories, {\it the particle passed through the holographic screen of the second trajectory, at some time in the past.}  This conclusion depends on the localization of the particle state in space-time.
The notion of particle track used in this argument is intuitive and familiar, but needs justification in the HST context.  We will provide this in the Appendix.

The entire class of Hamiltonians above preserves the particle/horizon separation of asymptotic states.  Indeed, the Hamiltonian can only change an N independent number of the constraints, because the polynomial has fixed order\footnote{In higher dimensions, the order will grow like $N^{d-4}$, but this cannot eliminate $N^{d-3}$ constraints.}, and because it acts on only a few of the DOF at short times, and its action goes to zero at large times.   Note that we have not specified the action of $H_{out}$, but as we suggested above, this will be determined by consistency of this trajectory's description, of physics that is described by the $H_{in}$ of other trajectories.   The constraints will be important for the particle DOF, and will lead to a picture of the interactions in terms of time ordered Feynman diagrams.

In fact, although most of the Hamiltonians we have written, will not satisfy the constraint that the S matrix is Lorentz invariant, they will {\it all} have a notion of Feynman diagrams for particles, or rather the time ordered diagrams appropriate to a Hamiltonian formalism in physical gauge.  Consider a subset of the incoming asymptotic particle DOF, which, following the argument in the appendix, continue to be incorporated in the $H_{in} (n)$ of some particular geodesic, $T_1$.  As long as $n \gg \sum K_i$, we can follow the tracks of these particles to smaller and smaller diamonds.  Once this inequality is violated, at some $n_{crit}$ the particles lose their identity, and the 
separation between particle and horizon DOF breaks down.  As we proceed through this small diamond to larger diamonds in the future, the Hilbert space of $H_{in} (n)$ eventually satisfies the inequality again.  As long as nearby trajectories have not enforced any constraints corresponding to more incoming particles, we will again have a new set of $n \sum K_i + q$ constraints, which enable us to identify the outgoing states as particles.  The causal diamond in which the particles lose their identity acts as a non-local vertex, which converts some set of incoming to some set of outgoing particles\footnote{Of course, one possibility is that the outgoing state is just a generic state in the ``in" Hilbert space at $n_{crit}$, which continues to evolve under the interaction terms of the Hamiltonian, on a time scale of order $n_{crit}^{-1}$.  This corresponds to black hole formation if $n_{crit} \gg 1$.
The probability of spontaneously producing a particle state under this evolution is
$e^{- n_{crit} E}$, where $E$ is the particle energy, so such a black hole will decay at the Hawking rate. The time scale for the total decay process is $\gg n_{crit}$ so this will not be interpretable as a local vertex for particle production.}.  The non-locality is on the scale $n_{crit}$, if we ignore the black hole formation processes discussed in the previous footnote.

\begin{figure}[!htbp]
\begin{center}
  \includegraphics[width=9cm]{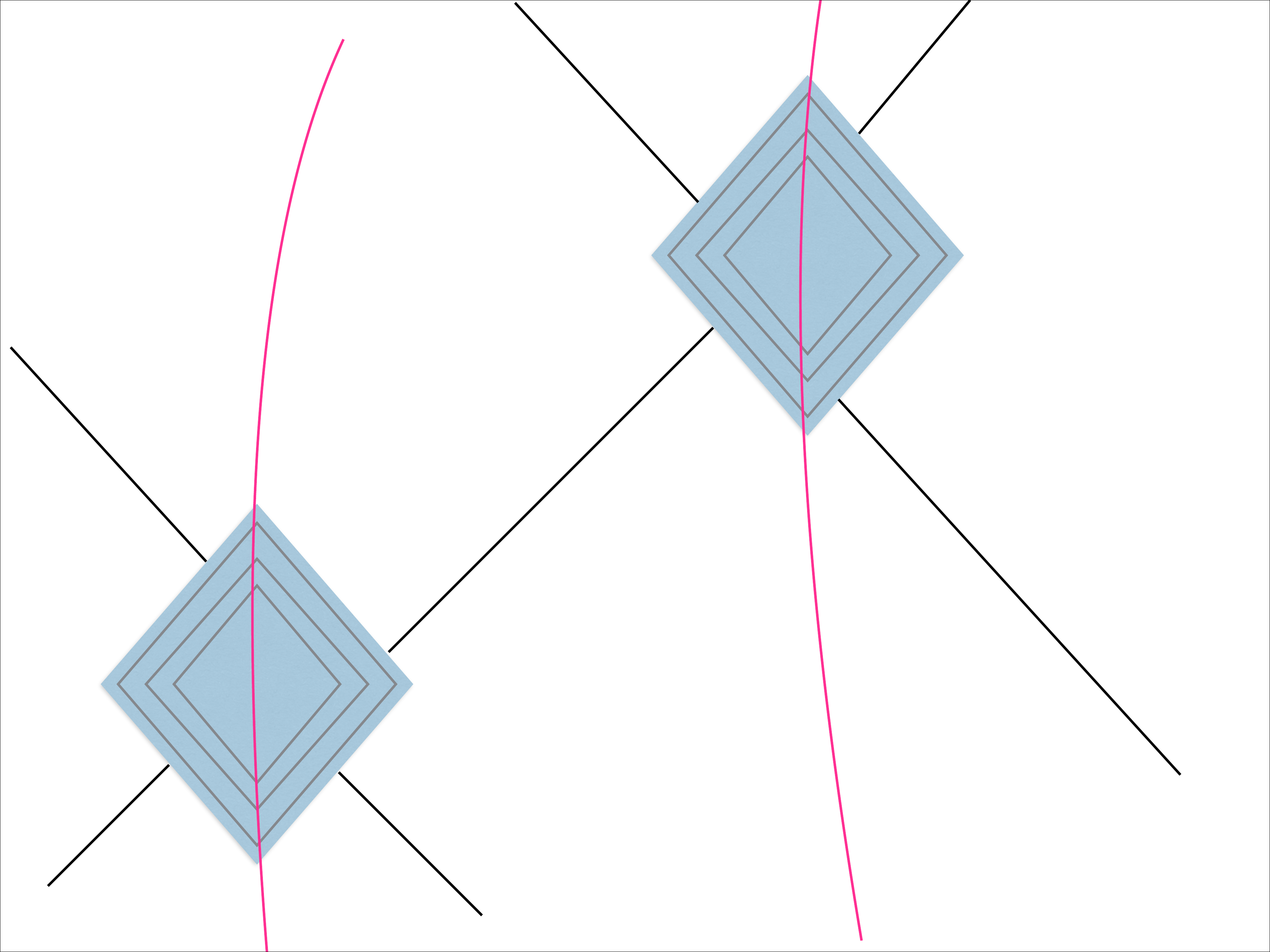}
  \end{center}
\caption{Fig. 1 A Time Ordered Feynman Diagram Describing Part of the Action of the HST Hamiltonian}
\end{figure}
Now consider another trajectory, $T_2$, whose point of time symmetry is chosen to be at some later time than the causal diamond that gave rise to the vertex.  The overlap conditions guarantee that the Hilbert space of $H_{out} $ of $T_2$ contains a copy of the state of particles in the space of $H_{in}$ of $T_1$, and vice versa.
If a particle created at the $n_{crit}$ vertex enters into a small causal diamond of $T_2$, it can participate in a new interaction.  We can depict this as a time ordered Feynman diagram, as in Fig.1.  Thus, the HST Hamiltonians and overlaps, reproduce the pictures of particle interactions familiar from QFT.  The effective vertices are non-local, as anticipated from Wilsonian considerations, but (see previous footnote) if $n_{crit}$ is large enough, a new phenomenon occurs, in which the ``vertex" remains in equilibrium with itself for a very long time, slowly emitting particles.

In \cite{newton} we showed that the large impact parameter eikonal scattering (the massless particle version of Newton's Law), {\it also} followed from the HST formalism, though in a very different way.  It appears as a non-local effective interaction and is mediated by the active horizon DOF, which are absent in the QFT description of the system.  Indeed, in the QFT description of quantum gravity, the long range instantaneous interactions are put in by hand (though of course they can be derived by proper gauge fixing of the original covariant Lagrangian) to satisfy Lorentz invariance in physical gauges.  

The main role of the present paper is to describe black hole formation and evaporation processes which cannot be described in terms of such diagrams. Before proceeding to review and revise our HST model for black holes, we must address a question of principle.  In HST, the structure of space-time is encoded in non-fluctuating properties of a collection of quantum systems - the dimensions of the overlaps.  A black hole of finite entropy can be formed by a scattering process in Minkowski space.
In HST, the description of this process never, in principle, introduces a space-time structure different from Minkowski space.  The black hole is a quantum state in Holographic Minkowski space, rather than a different space-time geometry.  Indeed, in HST, we've just seen that there is a sense in which all particle interactions arise from formation and evaporation of mini-black holes, which are too small to come to complete equilbrium.  If too much energy enters the past boundary of a sufficiently small causal diamond, the incoming state uses up all of the DOF in the diamond, and within a light crossing time the particle states are completely scrambled and the final state is random.  As the randomly distributed energy emerges from the future boundaries of larger diamonds, it can be described in terms of particles again.  If the scrambled diamond is sufficiently small (smaller than the resolution of our experiment, to use effective field theory language, but also small enough that the state of the diamond does not have a  high probability to become a long lived meta-stable equilibrium), then the process can be encoded as a small set of amplitudes for scattering of particles.  If it is very large in Planck units, it will form a long lived meta-stable state, which is what we call a black hole.

Jacobson's principle tells us that the space-time metric of the black hole is a hydrodynamic description of the collective behavior of the large number of DOF making up this meta-stable state.   The notion of trajectories that fall into the black hole after it forms and before it evaporates completely, is an emergent one.   We will find below that our description of the black hole from the external observer's point of view, contains a short lived image of the interior of the black hole, whenever particles fall into it.  This is, in some sense, a quantum version of the membrane paradigm, in which features of the interior are discernible on the horizon for a scrambling time (see Appendix B).   Trajectories in the interior, are simply described by prescribing overlap rules for causal diamonds ({\it i.e.} tensor factors), inside the Hilbert space of the external trajectory.

\subsection{The Black Hole Exterior}
\begin{figure}[!]
\begin{center}
  \includegraphics[trim = 30mm 40mm 30mm 20mm, clip, width=12cm]{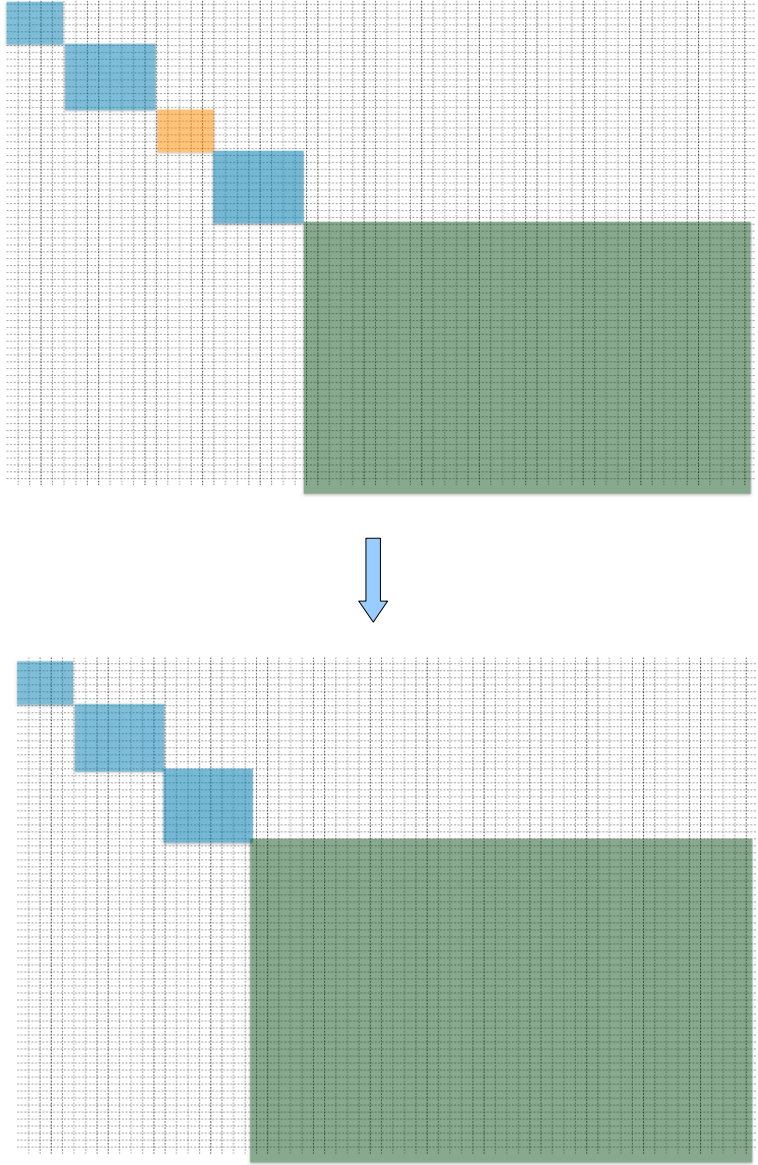}
  \end{center}
\caption{The matrix configurations corresponding to the time evolution of a particle (orange block) falling into a black-hole and eventually  becoming part of the active horizon states (large green block). Note that these matrices should be thought as ``compact" as the physics is invariant under permutation of blocks.}
\end{figure}
The fundamental description of a black hole in HST is given by a set of Hamiltonians $$H = H_{in} (n, {\bf x}) + H_{out} (n, {\bf x}) ,$$ with $ 1 \leq n \leq N$, describing physics from the point of view of a set of geodesic trajectories in Minkowski space.  The trajectories are at rest w.r.t. each other, and are parametrized by a cubic lattice ${\bf x}$ with lattice spacing equal to the Planck length.  The Hamiltonian $H_{in}$ is built from a set of variables $\psi_i^a (P)$, which are $n \times n + 1$ matrices, extracted from $N \times N + 1$ matrices $\psi_I^A (P)$.  $H_{out}$ is a function of the remaining matrix elements of  the $N \times N + 1$ matrices.  There is a separate set of variables for each ${\bf x}$.  The overlap rules are partially specified by insisting that the overlap Hilbert space at time $n$, for two points which are separated by $d$ lattice steps is spanned by the action of $n - d \times n - d + 1$ generators.   The Hamiltonian
$H_{in} (n)$ is the same at every ${\bf x}$ and has the form
$$H_{in} (n) = \sum P_0^k + \frac{1}{n^2} \sum g_p (n) {\rm tr}\ (\psi \psi^{\dagger} )^p .$$
Although this looks ``the same" for all ${\bf x}$, there are actually differences, which depend on the definition of particles.   Recall that particles are defined in terms of asymptotic boundary conditions at time $\pm N$ (with $N \rightarrow\infty$ eventually).  The particle states are well defined wave packets in the following sense: 
{\it for each {\bf x}, some of the particles will remain particle states down to some small $n_{min}$ .}  That is, the incoming state at time $- n_{min}$ and position ${\bf x}$ still obeys the constraints that distinguish particles from horizon DOF, for some subset of the incoming particles in the large (size $N$) causal diamond.

Other particle states will not be included in the DOF described by $H_{in} (- n_{min}, {\bf x})$, but might still look like particles at some other lattice point, at time $ - n_{min}$.  These states ${\it will}$ be described by $H_{out} (- n_{min}, {\bf x})$, as long as $N$ is large enough that the $[-N, N]$ causal diamond contains the causal diamond where the interaction occurs.  The overlap conditions on the density matrix will be constraints, which determine $H_{out} (- n_{min} , {\bf x})$, as well as the particular sub-factor of the Hilbert space at ${\bf x}$, which describes information shared with ${\bf y}$.  We have not yet worked out the form of these constraints, but we have argued above that one consequence of these is that the particle S-matrix is invariant under spatial translation.  The simplest hypothesis is that $H_{out} ( - n_{min} {\bf x})$ simply contains a copy of
the DOF which are described as particles in $H_{in} (- n_{min} , {\bf y} )$, with the same Hamiltonian.  This is surely a good approximate description if the physical distance between the points is large enough.

We can also describe the space-time in terms of any another set of time-like geodesics, boosted to an arbitrary $3$-velocity w.r.t. our canonical set.  Again, because the particle sub-factors of the Hilbert spaces describing asymptotic causal diamonds have complete overlap, the overlap constraints imply that there is a unitary representation of the full Poincare group on each geodesic Hilbert space, in the limit $N \rightarrow\infty$. The Poincare algebra commutes with the S-matrix.  The definition of momentum in terms of super-charges implies that this will in fact be a unitary representation of the super-Poincare algebra.  The only one of these asymptotic conservation laws that is realized by every member of the class of Hamiltonians we exhibited above, is energy conservation, which is the quantum number we labelled by $K$.  The other super-Poincare constraints should help to determine the couplings $g_p (n)$.  Finally, we note that time reflection (TCP) invariance is automatically incorporated by our requirement $H(n) = H(-n)$.  

From this fundamental point of view, we never have to mention the black hole interior.  The black hole is just a meta-stable resonance in the scattering amplitudes defined by our Hamiltonian.  We want to emphasize that, although the overlap conditions of HST require that we choose the coefficients $g_k (n)$ in such a ways that the S-matrix is super-Poincare invariant, every one of the Hamiltonians we study has primitive notions of particles and black holes.  Indeed, in a recent paper\cite{newton} we showed that they all gave rise to large impact parameter scattering amplitudes, whose scaling behavior with energy and impact parameter obeyed Newton's law of gravitation.

During the time that the black hole exists, the meta-stable excitation sequesters of order $R_S^2 (t)$ DOF from the rest, many of which  remain sequestered for times of order $R_S^3 (t)$.  For the external trajectories, all of these DOF are localized on the stretched horizon of the black hole.  This geometrical fact is expressed through different overlap conditions for different external trajectories. Thus, we can break the Hamiltonian of an external trajectory up into {\it three} commuting pieces $H(t) = H_{BH} (t) + H_{in} (t) + H_{out} (t) $, where the first term involves only the sequestered DOF.  As the black hole evaporates, DOF are transferred from $H_{BH} (t)$ to either $H_{in}$ or $H_{out}$ depending on the causal relation between the black hole and the trajectory in question.  Eventually, all DOF get transferred to $H_{in}$, except for the $N\rightarrow\infty$ limit of the active horizon DOF (the large block).  The Hamiltonian $H_{BH}(t)$ describes a system in an adiabatically varying equilibrium state.  The overlap conditions are such that, for time scales $\ll R_S^3 (t)$,  the Hilbert space on which $H_{BH} (t)$ acts, has no overlap with the Hilbert spaces on which $H_{in\ (out)}$ act.  For the external trajectories, this is simply a dynamical consequence of the fact that the Hamilonian we have used to describe scattering keeps the black hole at thermal equilibrium. The Hamiltonian of the black hole is

$$H_{BH} (t) = R_S (t) + \frac{1}{R_S^2 (t)} {\rm tr}\ g_p M^p .$$  $R_S (t)$ varies adiabatically, according to a process we will describe in a second, on a time scale $\gg R_S $.   Thus, to a good approximation, this is a time independent Hamiltonian
which has a natural time scale $R_S$, according to 't Hooft scaling.   In addition, because the trace of polynomials mixes up all matrix elements of a matrix, it is natural to conjecture\cite{sekinosusskind} that this Hamiltonian is a fast scrambler, with scrambling time $\sim R_S {\rm ln}\ R_S$.  Note that for our purposes it is not necessary that ${\it all}$ Hamiltonians of the form above are fast scramblers.  We only need the fast scrambling property to be valid for enough values of $g_p$ that we can also satisfy the constraints of Lorentz invariance.

A particle of energy $E$, inside the black hole, is a state annihilated by $o(E R_S)$ of the $R_S \times R_S + 1$ fermionic operators $\psi_i^A$.  A basis of such states are tensor product states of the non-zero DOF.  The probability of finding such a state in the maximally uncertain density matrix is $e^{- E R_S}$, which is a thermal probability at the Hawking temperature.   Once the off diagonal operators vanish, the particle has a chance of escaping the remaining DOF in the black hole.  We have not attempted to model the gray body factors, which determine which particles actually manage to escape the black hole.

Depending on the initial scattering state, particles may fall into the black hole at some time $t_h$ and angular direction ${\bf \Omega}$.  We always take $t_h$ to be at least a scrambling time after black hole formation.  In HST, as we have seen, the particle tracks are identified by following the pattern of constraints that are turned on and off by the action of the Hamiltonian as we follow the trajectory. 
A particle entering the black hole simply means that the particle DOF, as well as the off diagonal frozen horizon DOF, which connect the particle and the black hole, are added to $H_{BH} (t)$, because the action of the Hamiltonian is about to unfreeze the off diagonal DOF and equilibrate the particle DOF with those of the black hole.
The Hamiltonian now takes the form
$$H(t) = (H_{BH} (t_h) + \sum P_0^i) + H_{in}^{\prime} (t_h) + H_{out} (t_h) .$$ In writing this Hamiltonian, we have assumed that the black hole was in causal contact with the trajectory whose Hamiltonian we are modeling.  The ${\prime}$ on $H_{in}^{\prime} (t_h)$, indicates that the particle DOF, which were, prior to $t_h$, included in this Hamiltonian, are now in $H_{BH} $.  The black hole Hamiltonian 
after $t = t_h$ now includes the entire matrix formed by the original black hole DOF, the particle DOF, and the off diagonal terms, which couple them.  The interaction term has a factor $\frac{1}{(R_S + E)^2},$ where $E$ is the incoming particle energy.  The initial state, just after horizon crossing, has these off diagonal DOF zero and is a tensor product of a particle state and a black hole state.   Note that the state of particles falling into the black hole is {\it not} a generic state of the newly enlarged black hole, and will not become one until a scrambling time has passed.  This is true no matter how old the black hole is.
The act of dropping something into a black hole has long been known to increase the entropy of the hole by a large amount, which is still a small fraction of the total.
The HST model makes explicit the fact that the state of this higher entropy system is not generic, for a time of order the scrambling time.  {\it Thus, Page's ansatz that the black hole state is generic after a scrambling time, is violated each time something falls into the black hole, for a time of order the new scrambling time.  This is in accord with classical GR, which says that perturbations of the black hole leave a discernible imprint on the horizon, for a scrambling time.}
This feature of classical GR should be reproduced by any quantum theory that purports to reproduce this classical phenomenon as part of its hydrodynamics.  The model of black hole quantum states used by AMPS, where a finite fraction of the entropy is described by states of a free field in Schwarzschild coordinates, {\it does not} have a hydrodynamical description that behaves like classical GR.  Indeed, in the in-falling observer's frame it does not have any entropy and consists of a pure quantum state.  Entanglement entropy is not thermodynamic entropy.

Our description of the black hole from the point of view of a trajectory that never falls into it, appears to contain more information about the interior than the classical membrane paradigm.  It is known that the membrane paradigm appears to describe events inside the horizon, if they were determined by initial conditions outside the horizon.  We provide an example of this in Appendix B. However, in the classical membrane paradigm, particles falling through the horizon leave a temporary imprint on the horizon, but there is no way to encode the amplitudes for particles to interact behind the horizon in the
classical fields on the horizon.  On the other hand, in our model, the Hamiltonian $H_{BH} (t)$ does describe particle interactions inside the horizon, if they occur on a time scale $\ll R_S (t)$.  

We believe that this is a physically reasonable quantum extension of the membrane paradigm.  Consider an $e^+ e^- \rightarrow \mu^+ \mu^- $ amplitude, with the electron and positron momenta in the plane transverse to some Rindler horizon.  The Feynman diagram for this event contains contributions from space-time configurations where the virtual photon crosses the horizon and the muon pair is created outside of the causal patch of a Rindler observer.  However, there is no invariant way to separate out this contribution from the rest of the amplitude.  From the Rindler observer's point of view, there is an amplitude for a charge density with separated charge to appear on the horizon, accompanied by the gravitational field of a pair of separating muons.  We do not know of any attempt to generalize the membrane paradigm in perturbative field theory to discuss such quantum processes.  HST provides a proposal for doing so in a full theory of quantum gravity.

\subsection{The Black Hole Interior}

Jacobson\cite{ted} taught us that space-time geometry is a description of the hydrodynamics of quantum systems obeying the area/entropy rule, as HST does. Thus, we should view the distortion of Minkowski geometry created by a large black hole as a coarse grained, emergent description of certain meta-stable states in the underlying quantum theory.  In a similar manner, ``time-like trajectories that cross the black hole horizon" is a phrase describing emergent physics in this geometry.  The essential point is one of time scales.  There are at least three large time scales involved in black hole physics for a black hole of horizon radius $R_S$, the thermalization time $R_S$, the scrambling time $R_S {\rm ln}\ (R_S) $, and the evaporation time $R_S^3$, which is also of order the Page time.  The emergent trajectories for a large black hole have the properties of the trajectory of a geodesic observer
in Minkowski space before the black hole forms and after a time of order $R_S^3$ post formation.  At times intermediate between formation and evaporation, these trajectories behave differently, and one of our tasks is to construct a consistent quantum description of them.  All questions about firewalls and information loss have to do with the behavior of the Hamiltonian and Hilbert space of the trajectory, during this time period.  We will discuss a spherically symmetric black hole centered around a point that we will call the origin of our lattice of trajectories.  The black hole forms at time $t=0$, and has evaporated at $R \sim R_S^3$.  At intermediate times it can be viewed as a Schwarzschild metric with a time dependent $R_S (t)$, whose variation is determined by the Hawking radiation law.  It is very slow on the time scale $R_S$ and this variation is completely analogous to the adiabatic variation of equilibrium states in contact with an environment.

We will discuss trajectories that are the analog of Novikov coordinates, which are parametrized by in-falling radial time-like geodesics, at rest at $t=0$ in Schwarzschild coordinates.  The Novikov trajectories are parametrized by an angle, and a time $t_h$ at which the trajectory crosses the horizon of the static black hole.  Spherical symmetry implies that physics does not depend on the angular position of the trajectory on the black hole horizon.

Of course, we are talking about trajectories for a black hole formed in collapse, rather than the eternal Schwarzschild metric, so we really mean coordinates analogous to Novikov coordinates for that metric, radial time-like geodesics that fall through the black hole horizon some time after it forms.  In all of what follows, we will work to leading order in the ratio of black hole mass to the Planck mass.  Some of the phenomena we will discuss are non-perturbative in that ratio, but what we mean is that we capture just the leading order description of the phenomenon, neglecting irrelevant higher curvature terms.  

One other feature that distinguishes our trajectories from the fixed radius Schwarzschild geodesics that define Novikov coordinates is the fact that we take black hole evaporation into account.  Thus, each trajectory is parametrized by an angle, a time $t_h$ at which the trajectory ``falls through the horizon" and a horizon radius $R_S (t_h)$, which shrinks slowly.  We will study three trajectories: an early falling trajectory, for which $t_h = t_E$ is about one scrambling time past the time of formation of the horizon, a late falling trajectory, for which $R_S (t_L) < \frac{1}{\sqrt{2}} R_S$ (so $t_L$ is later than the Page time) and an external trajectory, such that $R_S (t_{ext} )\sim 1$.  
In other words, the external trajectory never falls into the black hole at all, since it has evaporated before that trajectory gets to the origin.  This is the trajectory discussed in the previous section.

 Our time parameter along internal trajectories is synchronized to that of the external trajectory.  It will coincide with proper time along each of the other trajectories, until that trajectory hits the singularity and the geometrical picture of the interior of the black hole breaks down, as seen from that trajectory. Proper time does not however end at that point, and we no longer believe that the time dependent Hamiltonian becomes singular or has Planck scale time dependence as the singularity is approached.   
 
 In fact, we will claim that all in-falling trajectories have, at least in the low curvature approximation in which we are working, exactly the same sequence of time dependent Hamiltonians as the external trajectory.  Different trajectories differ only in their overlap rules.   Indeed, the description of the previous section included the entire history of formation and evaporation of the black hole, including the effect of in-falling particles.   We also noted that that description included the amplitudes for particles scattering inside the horizon, before they hit the singularity, and interpreted this as a quantum extension of the membrane paradigm.
 
 Consider a time-like geodesic that falls into a black hole of initial Schwarzschild radius $R_S$ at a time $t_h > R_S {\rm ln}\ R_S$ after it forms, and at an angle ${\bf \Omega}_1 $.  We use the same sequence of time dependent Hamiltonians
 $$H(t)  = H_{BH} (t) + H_{in} (t) + H_{out} (t) .$$ that we used for the trajectory that ``falls in" only after the black hole has evaporated.  The time dependence of $H_{BH} (t)$ includes both the effects of Hawking evaporation, and growth of the black hole when it is hit by particles.   This sequence of operators all operate in a Hilbert space ${\cal H}_{BH}$of some maximal dimension, which is determined by the maximal horizon area attained in the course of black hole formation, in-fall of extra particles, and evaporation.
 
 To distinguish our particular in-falling trajectory, we introduce, for $t > t_h$,  a tensor factorization of the operator algebra on ${\cal H}_{BH}$ into the algebra of operators accessible in the diamond formed by the segment of trajectory between
 $t_h$, and $t$, and its commutant in the algebra of all operators on the space.
 $$B({\cal H}_{BH}) = B_D^1 (t) \otimes B_D^{1\ \prime} (t) \footnote{$B_D^{1\ \prime}$ is the commutant of the diamond algebra $B_D^1 (t)$ in the algebra of all operators 
 $B( {\cal H}_{BH})$ on the black hole Hilbert space.} .$$
 We now use the semi-classical geometry\footnote{By semi-classical, we mean that we take into account the shrinkage of the black hole due to Hawking radiation.} to assess whether any of the particles that fall into the black hole are in causal contact with our trajectory at any time between $t_h$ and the time that the trajectory hits the singularity.  If particles are in causal contact at time $t$, then the operators describing them are included in the algebra $B_D^1 (t)$.   Recalling that particles are, by definition in HST, localized in angle, then at $ t \sim t_h$ most particles will not be causally connected to our trajectory.  However, the shrinking of two spheres after horizon crossing, in the semi-classical geometry, tells us that particles which fell in during a time interval $\sim R_S$ bracketing $t_h$ will come into causal contact by the time the singularity is hit.
 
 Consider another trajectory, which falls into the hole at the same $t_h$ but at a different angle ${\bf \Omega}_2$,  Then the overlap between $B_D^1 (t)$ and $B_D^2 (t)$ will be empty for $t \sim t_h$.  However, when we reach the singularity,  which occurs at $t \sim t_h + R_S$, the particle DOF begin to scramble with the horizon DOF and by $t_S \sim t_h + R_S {\rm ln}\ R_S$, the entire system is in equilibrium and accesses the entire Hilbert space.   At this point $B_D^1 = B_D^2 = B({\cal H}_{BH}) $.   The complete overlap of these operator algebras is the signal of the shrinking of transverse two spheres to zero size in the black hole interior.  It is not a singular phenomenon in the quantum theory along any trajectory.
 
 The other singular aspect of the black hole interior geometry is the way in which space ``recedes faster than the speed of light in the radial direction, achieving infinite size in finite proper time".  The area near the horizon always remains non-singular, with curvature of order $R_S^{-2}$.  In our quantum description this is a consequence of the fact that the external observer can see particles falling into the black hole at times much later than the scrambling time.  We can model the physics along trajectories that fall in, in such a way that they are in causal contact with these new in-falling particles, in exactly the same way that we did above, with the same results:  ordinary particle physics for $t_h^{\prime}   < t \ll t_h^{\prime} + R_S (t) $, followed by thermalization.  The overlap conditions of the semi-classical geometry tell us that there is no overlap between the events of particle scattering at this later time, and particle physics that happened inside the horizon at times of order $t_h \ll t_h^{\prime}$.  This is of course easy to implement in QM, and is the signature of the expansion of the geometry in the radial direction.  From the point of view of the Hamiltonian, it is simply the statement that the phrase ``particles falling through the horizon" corresponds to the enlargement of ${\cal H}_{BH}$ by an entropy of order $E_{particle} R_S$, coupled with a constraint on the initial state in this larger Hilbert space.  If this occurs at $t_h^{\prime} \gg t_h$ then there is no surprise that the factor algebras corresponding to causal diamonds between $t_h $ and $t \leq t_h + R_S {\rm ln}\ R_S$ have no overlap with the diamonds describing particle physics after $t_h^{\prime}$.   That physics involves new DOF which were brought in with the in-falling particles. Prior to that time, those DOF were not a part of the algebra $B({\cal H}_{BH}) (t) $.  
 
 In the actual QM of the system, the DOF encountered by the trajectory that fell in at $t_h$, are a subset of the algebra of active horizon DOF of the black hole.  They indeed commute with the new operators that we have added to describe particles that fall in around the time $t_h^{\prime}$.   
 Again, the QM is non-singular, and the space-time picture is singular because it fits the quantum mechanically obvious tensor factorization of the operator algebra described in the preceding paragraph into a framework in which commutation of operators is attributed to space-like separation.
 
 Notice that there is no sign of a firewall in these considerations, and $t_h^{\prime}$ could easily be taken larger than the Page time of the originally formed black hole.
 The AMPS argument fails for multiple reasons in HST.  It is certainly true that the original black hole Hilbert space comes into thermal equilibrium and that if we wait a Page time the outgoing Hawking radiation will be maximally entangled with the remaining states of the black hole.  However, as we have seen, in HST the description of particle physics inside the horizon requires us to take into account the enlargement of the black hole Hilbert space due to the energy of the in-falling particles.  This enlargement is very large in absolute terms, unless the total energy is about that of a single Hawking quantum (which means that it can't really be localized inside the horizon). Furthermore, the initial state in this enlarged Hilbert space is not generic, and most of the added entropy is frozen.  It takes a time of order the Schwarzschild radius for equilibrium to begin to set in for the enlarged Hilbert space.  The physics of trajectories discussed by AMPS, which fall into the black hole at times larger than the original Page time, yet according to classical geometry manage to see some particle physics events before hitting the singularity, is precisely what we discussed in the previous paragraph.  In HST this physics has to do with DOF that do not form part of the original black hole equilibrium state, and are not entangled with the early Hawking radiation.
 
 The reader who has followed our previous work\cite{fw1} will note that our story has changed dramatically, but that our conclusion has not.  {\it There is no firewall.} All descriptions of black hole formation and evaporation in HST have the property that in-falling particles remain decoupled from the black hole horizon for a time of order $R_S$.  Here we've argued for this using the fact that the black hole collision with the particles is modeled by a system where the off diagonal matrix elements connecting particles and black holes are equal to zero just before the particles cross the stretched horizon.  It would be perverse to make a discontinuous jump in the state of the system rather than have it evolve naturally to equilibrium by simply modifying the Hamiltonian so that the degrees of freedom that are incorporated into the black hole, at any time, are simply decoupled, for a time of order the evaporation time, from the rest of the DOF in the universe.  This is precisely the quantum statement of the fact that these states remain causally disconnected from the exterior for a period of order $R_S^3$. 
 There is no reason to modify this description after the scrambling time, the Page time, or any other time except that at which $R_S$ becomes too small for a thermodynamic description of the hole to make sense.   {\it Page's argument is never relevant to this discussion because a particle state {\bf in any causal diamond} in HST is, by definition, a state in which a huge number of the DOF are frozen and the small tensor factor of the remainder is not entangled with the bulk of the horizon DOF.  }   A particle state is not a generic state of any causal diamond, whether inside a black hole or not.   None of the consistency conditions of HST require entanglement between the radiation emitted from a black hole before the Page time and the particle states inside the hole.
 
 The Hawking radiation occurs as a consequence of the interaction of the horizon degrees of freedom of the hole, and the external DOF.   We have not presented a detailed model of the quantum mechanics of Hawking radiation, but our eventual model will have nothing to do with entangled Hawking pairs because no such field theoretic objects appear in the HST model.  The equilibrium state of the hole will produce a particle state of energy $E$, with probability (to leading order in $R_S$) $e^{- E R_S}$, so the system is thermal with the correct Hawking temperature.  This is a consequence of the fact that the definition of the particle state sets $o(R_S E)$ fermions to zero.  Thus, our model black hole will radiate particles at the correct rate, as must any model with the correct energy/entropy relation.   
 
 The free field theoretic model of Hawking radiation is simply inconsistent, because, in the frame of a geodesic observer it does not have the correct degeneracy of black hole states.   Instead it has only one low energy state, the vacuum.  As has been emphasized by Marolf\cite{marolffuzzorfire}, this is the essence of the AMPS paradox.  Indeed, the robust field theory argument that a black hole is a thermal system with the Hawking temperature, comes from the Euclidean calculation, which is applicable to arbitrary interacting field theories and suffers no UV ambiguities or cutoff dependence.  The naive picture of Hawking pairs is, we believe, simply incorrect. Jacobson's connection between Einstein's equations and hydrodynamics provides the ultimate explanation for why the classical black hole metric describes a thermal system.   In high entropy situations, hydrodynamics is still a valid classical theory, but quantized hydrodynamics is only a valid approximation to quantum systems near non-degenerate ground states.  The near horizon region of a black hole is not such a system.

 \section{The S-matrix From Emergent Trajectories}
\begin{figure}[!htbp]
\begin{center}
  \includegraphics[width=9.5cm]{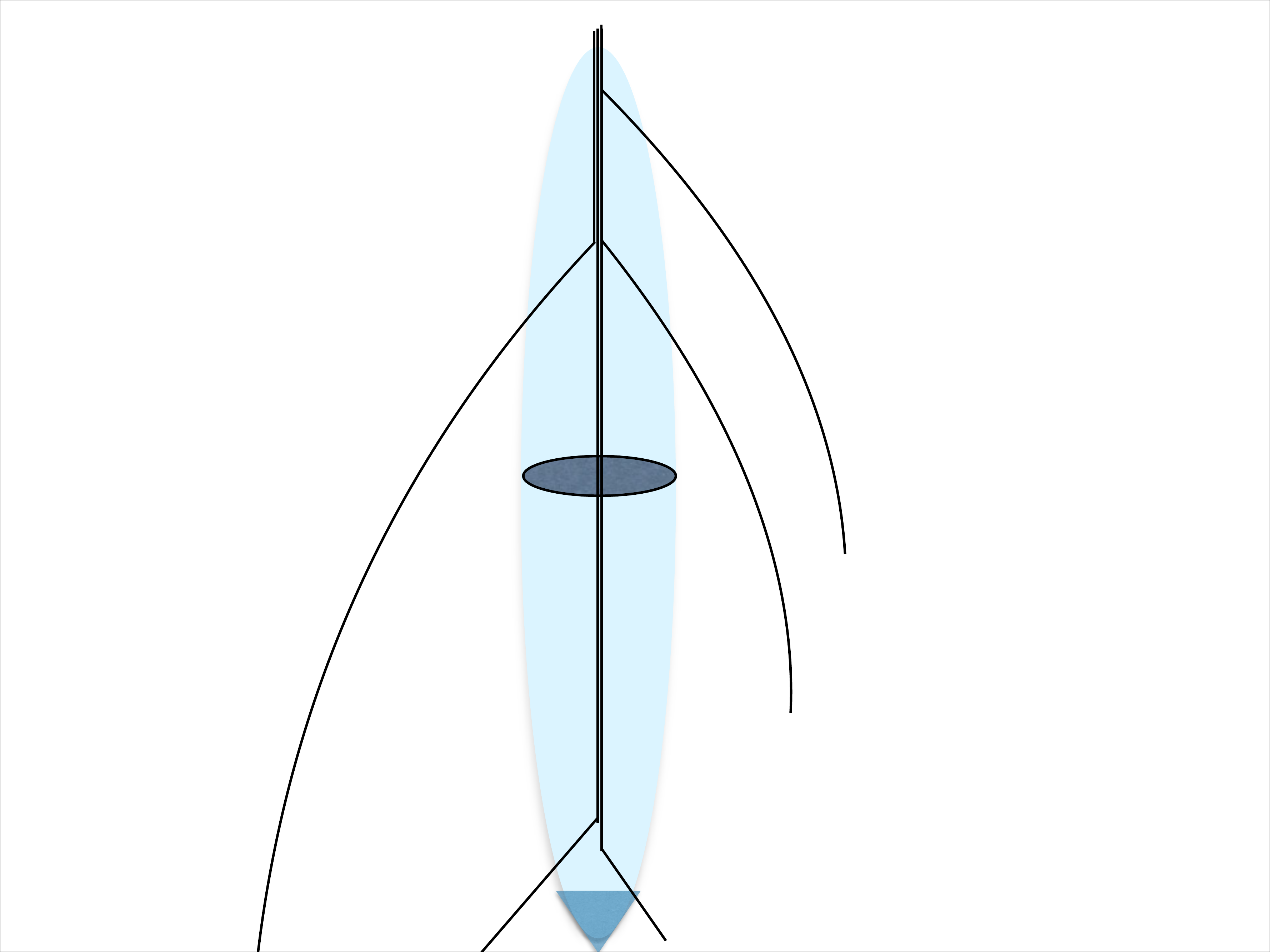}
  \end{center}
\caption{Particles falling into a black hole formed by collapse which subsequently evaporates.}
\end{figure}
We can now summarize what we've learned about the interior of a black hole in HST in a picture, Fig. 3.  This figure depicts the stretched horizon of a black hole formed in collapse as emanating from a point (a region of Planck scale size), which is the tip of a light cone.  After a scrambling time we draw the stretched horizon as a time-like surface of slowly varying radius.   In principle we could incorporate growth of the black hole due to in-falling matter but we've assumed that that is small, and our picture shows only the shrinkage of the radius due to evaporation.  The stretched horizon disappears when the Schwarzschild radius becomes so small that the thermodynamic description of the black hole is grossly inadequate.  Since we're not sure when that is, we depict the future end of the stretched horizon cylinder as a smooth cap.  

The figure also shows three trajectories that explore the black hole interior during evaporation in addition to the two in falling trajectories during collapse.  Two fall in at the same time during evaporation, but at different angles.   In a time of order $R_S$ they become an effective single trajectory, since their overlap is complete.  Similarly a trajectory that falls in later, no matter how much later, has physics identical to the earlier trajectories.  It's an important subtlety of HST that although these systems have identical physics after a certain time, they are independent QM systems, related at most by overlap conditions (and even that isn't true for an early and sufficiently late falling trajectory). 

The picture also shows that last trajectory emerging from the future boundary of the stretched horizon, where it actually stands for any trajectory that entered the black hole. 
Since we have taken the time dependent Hamiltonian along any of these in-falling trajectories to be identical to that of the external trajectory, the S-matrix computed along any of these trajectories is the same.  The only difference between them is the overlap conditions imposed on ``the causal diamond of the trajectory between horizon crossing and a time $t$ before the trajectory hits the singularity".  These conditions have non-trivial content only for trajectories that are in causal contact with some of the particles that fall into the black hole after it was formed.

\section{Conclusions}
We conclude by summarizing our basic framework. 

\begin{itemize}

\item HST in Minkowski space assigns of order $N^2$ pixel variables, each of which has a finite dimensional representation space, to a causal diamond whose radius in Planck units is $N$.  Of these at most $N^{3/2}$ can be classified as particles\footnote{For particle states of fixed energy much less than that of a black hole of horizon area $N^2$, the entropy bound is much smaller.}.  Particles and their energy are defined sharply only in the asymptotic limit $N \rightarrow\infty$.  Most of the $N^2$ DOF are associated with the horizon.  The bulk is defined, for each time-like geodesic in Minkowski space, by a sequence of time dependent Hamiltonians $H_{in} (n) + H_{out} (n) $, $1 \leq n \leq N$, which determine propagation from the past boundary of a causal diamond at time $- n - 1$, to the past boundary at time $-n$, as well as the propagation from the future boundary at time $n$ to the future boundary at time $n + 1$.  $H_{in} (n) $ is built from of order $n^2$ DOF and $H_{out}$ from the rest.  All experiments done on that trajectory during the interval $ [-n,n]$ are described by $H_{in} (m)$ with $|m| \leq |n|$, while $H_{out}$ is fixed for consistency with the descriptions of distant processes along trajectories that are in causal contact with them.  The simplest way to ensure this consistency is to have $H_{out} (n, {\bf x})$ for a trajectory labeled ${\bf x}$, contain copies of $H_{in} (n, {\bf y})$ for a trajectory ${\bf y}$, which passes through the other scattering event.  When the points are sufficiently distant from each other, this is probably a good approximation, but we have not written down an complete solution of the HST overlap conditions for Minkowski space. We have argued that general Hamiltonians in our class will have particle scattering amplitudes that correspond to time ordered Feynman diagrams  (with localized, but not necessarily Lorentz covariant vertices).
An excitation that can be localized in the bulk is one whose evolution can be followed from times of order $\pm N$ down to times of order $n \ll N$.  This is clearly untrue for most of the DOF.   When the initial and final states in a causal diamond of size $n$ can both be described as particles, we can write the amplitude for the process as an effective time ordered Feynman diagram, with vertices non-local on the scale $n$.  However, if $n \gg 1$ and all the DOF come to equilibrium, then the equilibrium is meta-stable.  Our matrix models have the property that the probability of finding a particle state of energy $E$ in the equilibrium ensemble is $e^{- nE}$, which shows that the particle emission spectrum is thermal with a temperature that scales like the Hawking temperature.   Thermodynamics then tells us that the lifetime of the equilibrium state is $\sim n^3$.

\item Scattering occurs mostly in small causal diamonds where the distinction between particle and horizon DOF is lost.  This is the origin of approximate locality.  In a causal diamond of size $R_S$, the generic state has energy of order $R_S$ and the levels are extremely dense.   This is a black hole.  The classical geometry of the black hole is a description of the hydrodynamics of this high entropy quantum system.  It is an emergent concept, for large $R_S$.  Associated with this geometry are emergent trajectories.  For this paper, the most important emergent trajectories are lines of fixed angle in (generalized) Novikov coordinates.  They are parametrized by an angle (but the dynamics is rotation invariant) and a time $t_h$ at which they cross the horizon.  We concentrate on two particular trajectories, for one of which $t_h = t_1$ about one scrambling time after the horizon forms.  The other has $t_h = t_2$, of order the Page time.  We also use as a reference a third trajectory whose ``horizon crossing time", according to the original Schwarzschild metric, is longer than the black hole evaporation time.  For this trajectory, black hole formation and evaporation is simply an exotic scattering event.  For the early and late falling trajectories it is more dramatic.

\item We modeled the dynamics along any trajectory, which falls into the black hole when its horizon is macroscopic, by the same time dependent Hamiltonian
$$H(t) = H_{BH} (t) + H_{in} (t) + H_{out} (t) .$$
$$ H_{BH} (t) = \sum P_0^k + \frac{1}{[R_S + E]^2 (t)} \sum g_p (R_S(t) + E) {\rm tr}\ (\psi_i^a \psi^{\dagger\ j}_a )^p .$$  
The particle energy here is that of particles that can be causally accessed in the causal diamond of the trajectory inside the black hole , $E = \sum P_0^k$.  We must take into account the fact that the black hole has grown due to the addition of particles, in order to implement the $E R_S$ constraints that separate particles from horizon states inside the hole.  The initial state at horizon crossing satisfies these constraints and is a tensor product of a particle state and a scrambled state of the un-frozen horizon variables.  The action of the Hamiltonian, over time scales of order $R_S {\rm ln}\ R_S$, removes the constraint and scrambles the entire Hilbert space.  We emphasize that these events of particles falling into the black hole are programmed into the initial scattering state, which also determines the size and location of the black hole itself.

\item  In HST, the S-matrix can be computed using {\it any} of the trajectories we have discussed, independently of when (or if) the trajectory falls through the horizon, and of the angle.   The only difference between different in-falling trajectories consists in a tensor factorization of the operator algebra in the Hilbert space of the black hole into a factor describing the causal diamond of that trajectory 
between horizon crossing $t_h$ and some time $t$ with $t_h < t \ll R_S {\rm ln} R_S$.   The significance of this factorization has to do with when certain particle scattering processes behind the horizon come into causal contact with the trajectory in question.  The only question about this factorization that effects the dynamics is whether any such factorization is consistent with the form of the Hamiltonian.   If we ignore the fact that, at the moment, we only know how to describe massless particles in HST, and associate a massive detector with the trajectory, then the constraints defining the localized detector (analogous to those defining particles) would guarantee that this factorization was possible for times much less than the scrambling time.  

\item In essence, our proposal for the description of the black hole interior is to
define a sort of quantum version of the membrane paradigm.  That is,  the external trajectory has a Hamiltonian describing particles falling in to the black hole, which includes scattering of those particles inside the stretched horizon before they reach the singularity.  We take the same Hamiltonian to describe physics as seen along trajectories that enter into the black hole, modifying only the overlap conditions, in a way that depends on the trajectory.
All of the singular aspects of the interior geometry are expressed as perfectly finite overlap conditions relating different in-falling trajectories.

\item Since all ``Novikov" trajectories have the same time dependent Hamiltonian, whether or not they fall into the black hole, the amplitudes for processes to occur inside the horizon along a given trajectory in some finite proper time interval require a specification of more structure than the unitary operators $U(t,t^{\prime})$ on the full Hilbert space.  A given segment of a given trajectory corresponds to a particular tensor factorization of the operator algebra.   Thus, the amplitudes for local physical processes accessible in a particular causal diamond, are not implicit in the scattering matrix, but require more structure.   This is quite analogous to the situation in quantum field theory, where the S matrix does not specify the Green's functions of local fields.   Complementarity is valid in this model in the sense that, if one specifies a particular trajectory by making an appropriate tensor factorization of the black hole Hilbert space at each time, then one can evolve the operators in any fixed causal diamond to $t = \infty $ and find a set of operators with the same spectral properties and operator algebra, defined on the space of asymptotic states.
It is however, quite hopeless to try to do this without using the detailed picture of the interior provided by our model.  It is not just a question of the difficulty of the quantum computation required to reverse the time evolution, but of the various choices that have to be made to specify which causal diamond in the interior we are talking about, and whether we are discussing a geodesic or some randomly accelerated trajectory that ends on the tips of the diamond.
 
\end{itemize}

With regard to the last point, we want to clear up a misconception that we have heard repeatedly in discussions of the black hole information paradox.   Some researchers in the field talk about ``finding out what happened to the cat if a Schrodinger's cat experiment is performed inside a black hole".  We believe that this phrase is a mis-construction of what quantum mechanics actually says about macroscopic objects.

Consider a less exotic example in which we perform the experiment inside a hermetically sealed capsule in inter-galactic space.  The quantum mechanical prediction for the result of the experiment is that the wave function of the system is 
$$ \psi = \alpha | + \rangle \otimes | Live\ y \rangle + \beta | - \rangle \otimes | Dead\ y^{\prime} \rangle  ,$$ where $|\alpha^2 | + |\beta |^2 = 1$ and $y$ and $y^{\prime}$ are two of the exponentially large collection of microstates, which correspond to a fixed answer to the macroscopic question ``Is the cat alive?" .  Decoherence theory tells us that the quantum interference between the two states in this superposition is doubly exponentially small (exponential in the entropy of the cat).  

The standard interpretation of quantum mechanics is that this wave function predicts the probability of occurrence of the live or dead cat states.  That is, if we do the experiment many times $|\alpha^2 | $ and $| \beta |^2$ predict the frequency with which we will find each result.  The extreme smallness of the interference terms tells us that Bayes' conditional probability rule, from ordinary classical probability theory will apply to these predictions.   This is what allows us to say that some particular thing happened.   The decoherent histories analysis shows us that quantum predictions will obey the classical logical rules from this point on, with exquisite precision.  For example, the conditional probability given by the above wave function, to find the cat alive again, after we've found it dead, is zero for all physical purposes\footnote{This phrase means that the time scale for recurrence of a live cat is the same as that for the cat's constituent atoms to spontaneously form a live cat.  It's much longer than the age of the universe.}.

This, according to the rules of quantum mechanics, is what it means for ``something to happen".  ``Happening" means that quantum information is encoded in collective variables of macroscopic systems, to which the rules of classical logic and mechanics, apply with exquisite precision.   Now let us explode a nuclear bomb inside the space craft and wait many years.   Even assuming that the debris from the explosion does not interact with anything else, we eventually get to a situation in which all of the quantum information about the state of the system is encoded in phase correlations between the states of individual elementary particles.   

Unitarity guarantees that we can run this state backward, and reproduce the quantum state $ \psi $.  {\it It does not say ANYTHING about the result of the particular macroscopic experiment that we performed inside the capsule before the explosion.}  Classical information, which tells us about things ``happening", is not preserved in QM.   Things can {\it un-happen}.  Thus, even under the hypothesis of exact unitarity with zero information loss, what QM predicts is only the frequency with which,  if we had created exactly the same black hole with the same Schrodinger's cat experiment performed inside it, an infinite number of times, the in-falling experimenter found a live or dead cat before hitting the singularity.  

Our analysis shows that in the HST model, we can, in principle, retrieve the magnitudes and relative phase of the complex numbers $\alpha$ and $\beta$ from the S-matrix, once we have made careful specification of the causal diamond and the trajectory along which the cat experiment is done.  Nothing can ever tell us what happened in a particular run of the experiment in a single black hole.

We want to conclude with a comment about the validity of our models.  The key difficult feature of a proof that they are indeed models of quantum gravity is the demonstration that one can choose the coefficients $g_k (n)$ in such a way that the S-matrix becomes Poincare invariant.  However, even without doing that, it is clear that all of our models have particle-like excitations whose tracks can be followed through the space-time defined by the quantum mechanics of the model.   In \cite{newton} we've shown that the large impact parameter scattering of these particles is dominated by Newton's law.

Furthermore it's clear that these particles can collide to form large systems whose size is arbitrary (but controlled by the incoming particle energies) and which reach equilibrium on a time scale of order their size.  The energy/entropy relation of those systems is that of black holes.   The laws of thermodynamics, plus the fact that the only other localized excitations of the models are particles, guarantee that the decay rate of these meta-stable excitations will be governed by Hawking's law.  It seems reasonable to call these excitations HST Black Holes, and they exist for a broad range of choices of the coefficients $g_p (n)$.

In this paper, we've shown that, when the HST black hole is large, there is an alternative way of computing the S-matrix, which involves constructions that exhibit the properties of trajectories inside an HST black hole.  The overlap rules for these emergent trajectories have the geometric properties of the HST black hole interior.  Using these alternative rules we can see explicitly that unitarity of the S-matrix for HST black hole formation and evaporation is compatible with particle physics persisting along trajectories in the interior, for a time of order the Schwarzschild radius.  All of the different interior trajectories have identical Hamiltonians and their geometric differences, as well as the singular interior geometry of the black hole, are expressed through overlap rules.  It's clear then that the conventional picture of black hole formation and evaporation, with no drama at the horizon, cannot violate unitarity or any fundamental principle of quantum information theory.  

The firewall is not the right choice among the alternatives that AMPS proposed for resolving their paradox.   One possibility that much of the literature rejects is that it is quantum effective field theory that fails, and that is the path we are suggesting.  We don't understand why there is so much resistance to this idea - to the extent that some authors propose modifying the rules of quantum mechanics rather than giving up on quantum field theory. 

\appendix

\section{Appendix}

Our argument for Poincare invariance of the particle S-matrix depended on the idea that one could follow a particle track through space-time.  Since the bulk of space-time is, in some sense, an emergent concept in HST, one must provide a definition of the intuitive notion of track.  The purpose of this appendix is to do that.

Particle/jet states are defined asymptotically, in terms of a very large causal diamond of size $N$, which will eventually go to infinity.  
The defining equation is (we drop labels referring to the compact dimensions of space)
$$\psi_i^A | Jet \rangle = 0 , $$ for $NK_i + Q_i $ matrix elements, with $1 \ll K_i, Q_i \ll N$.  The defining 
commutation relations of the pixel variables $\psi_a^A$ are invariant under $U(N) \times U(N + 1)$ so we can always arrange the vanishing variables so that the matrix $M = \psi \psi^{\dagger} $ is block diagonal when applied to the state, with a $K_i \times K_i$ block.  Multiple blocks refer to multi-jet states.

In the $N \rightarrow\infty$ limit, the $\psi_i^A$ become general sections on the spinor bundle of the two sphere, while the $M$'s become the algebra of functions on the two sphere.  We will take the limit in such a way that the $\psi (\Omega) $ are operator valued measures.  In this limit, the non-zero elements in the $K_i$ block converge to a measure with support in a certain connected domain of the two sphere. The constraint
$$\psi_i^A | Jet \rangle = 0,$$ converges to the statement that this
connected domain is isolated from the connected domains defined by other small blocks, and from the bulk of the sphere, represented by the large block of size $N - \sum K_i $.  The small non-zero blocks correspond to flows of energy into regions on the two sphere, because our formula for the Hamiltonian associates energies with these blocks.  Terms in the Hamiltonian corresponding to the large block, all go to zero as $N\rightarrow\infty$.  These active horizon states are in some sense topological DOF, but only asymptotically.  In \cite{newton} we showed that they are essential to obtaining the correct scaling of Newton's law for large impact parameter scattering.  

Now begin with a state satisfying this constraint at time $-N$ and propagate it into the future, using the Hamiltonian of some particular time-like geodesic.  The Hamiltonian $H_{in} (n) $ for $ n = - N + p,$ with $p \ll n$ acts on fewer degrees of freedom, and the interaction term in it gets stronger as $p$ increases.  For some $p$ the degrees of freedom corresponding to one of the $K_i$ blocks may not be included in $H_{in} (- N + p) $.  However, as long as a block is included, and $p \ll N$, then it will still satisfy $o(K_i N)$ of the constraints placed on the asymptotic state.  This is a consequence of the way the Hamiltonian goes to zero for large $- N + p$ and the fact that the interaction polynomial in the Hamiltonian has $N$ independent order.   Thus, we can still think of the $K_i$ block as representing states localized in a small domain of the two sphere.  As $p$ gets larger the localization gets fuzzier because we have fewer spherical harmonics to play with.   Nonetheless, we can use this localization in successive causal diamonds to follow the track of a particle as long as $| - N + p | \gg K_{(p)}$, where $ K_{(p)}$ is the total particle energy contained in the diamond.  

The fact that a large number of constraints are involved in the definition of localization of particle tracks, and that these constraints are dynamically robust, shows that particle tracks are objective features of the initial state, not subject to quantum gravitational fluctuations.  Similar remarks can be made about outgoing particle tracks.

\section{The Membrane Paradigm and the Black Hole Interior}
\subsection{Two charges viewed in an accelerated frame }
Consider a fiducial observer hovering just outside a black hole. For such an observer, the near horizon geometry is a good approximation and the metric takes the Rindler form
\begin{equation}\label{rindler}
ds^2=-\rho^2 d\omega^2+d\rho^2+dy^2+dz^2 \,.
\end{equation}
This can be regarded as a portion of Minkowski space, formally known as the Rindler wedge. In particular, under the redefinitions
\begin{eqnarray}
t&=& \rho \sinh \omega\, ,\\
x&=& \rho \cosh \omega\, ,
\end{eqnarray}
one arrives to the more familiar metric
\be
ds^2=-dt^2+dx^2+dy^2+dz^2\,.
\ee
For the accelerated observers, who follow orbits of the time-like Killing vector $\xi=\partial_\omega$, there is a horizon at the edge of
the Rindler wedge, $x=|t|$, or equivalently, $\rho=0$. We will replace the mathematical horizon by the stretched horizon, a time-like hypersurface located at $\rho=\epsilon$, where $\epsilon$ is about one Planck length. Our goal is to study the electrodynamics of probe charges in this background and the possible imprints on the dynamics of the stretched horizon.

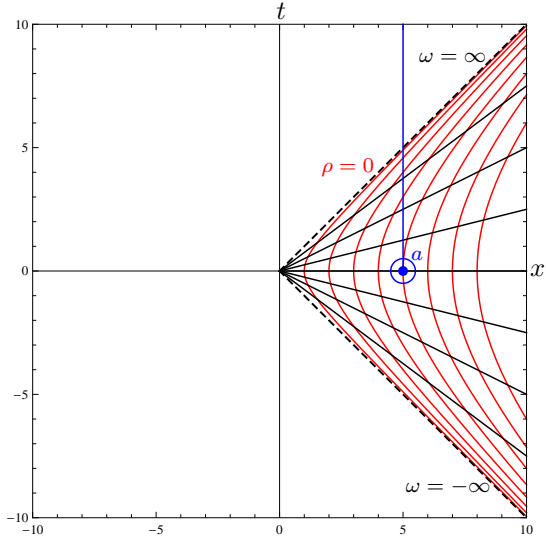
\begin{figure}[!htbp]
\begin{center}
  \includegraphics[width=7cm]{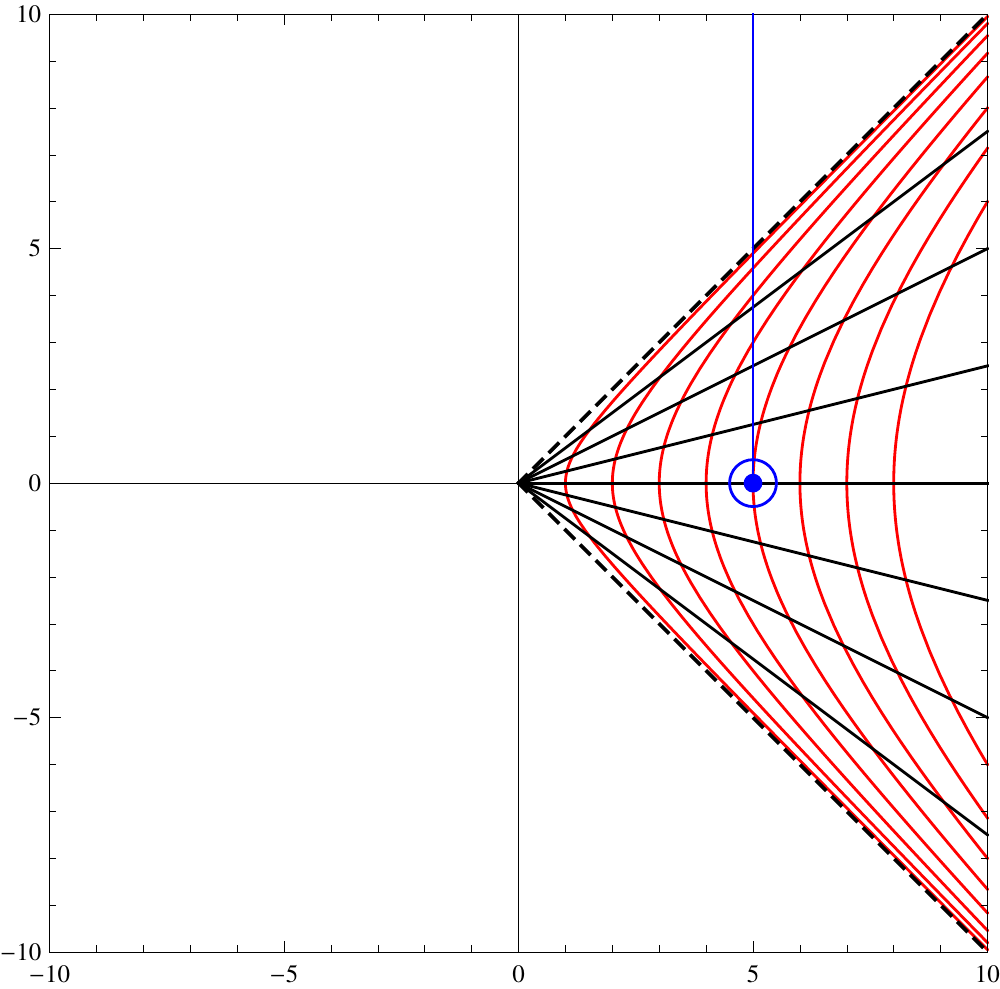}
   \begin{picture}(0,0)
 \put(-5,99){\footnotesize{$x$}}
  \put(-101,196){\footnotesize{$t$}}
  \put(-47,180){\scriptsize{$\omega=\infty$}}
  \put(-52,17){\scriptsize{$\omega=-\infty$}}
  \put(-84,138){\scriptsize{\color{red} $\rho=0$}}
  \put(-50,105){\scriptsize{\color{blue} $a$}}
\end{picture}
  \end{center}
\caption{Rindler chart plotted on a Minkowski diagram. The dashed lines correspond to the Rindler horizons. Constant-$\rho$ lines are depicted in red while constant-$\omega$ lines appear in black. The worldline for a free falling, time-dependent dipole (in the transverse plane) is plotted in blue.\label{fig1}}
\end{figure}

We are interested in a time-dependent dipole falling into the horizon \footnote{The discussion for the single charge case can be found in \cite{lindesay}}. We will model this configuration with two charges, one that is static and a second one that is moving (in the transverse plane) with constant velocity $v$:
\begin{align}
& \text{Charge}\ Q:\qquad x=a\ , y=\frac{b}{2}\ , z=0\ ,\\
&\text{Charge}\ -Q:\qquad x=a\ , y=-\frac{b}{2}+v t\ , z=0\ .
\end{align}
Figure \ref{fig1} represents schematically this situation.

We are interested in two particular cases: i) the two charges meet before crossing the horizon, and ii) the two charges meet after crossing the horizon. To compute the surface charge density induced on the stretched horizon, we need to determine the field component $E_\rho$. However, at any given time the Rindler coordinates are related to the Minkowski coordinates by a boost along the $x$-axis. Since the component of the electric field along the boost direction is invariant, we can write the standard Coulomb field,
\be
E_\rho=E_x\,.
\ee
Then, using Gauss's law we can infer the charge density induced on the stretched horizon,
\begin{equation}
 \sigma(\omega, y , z)= \frac{1}{4\pi \rho}E_x(t(\rho,\omega),x(\rho,\omega),y,z)|_{\rho=\epsilon}\,.
\end{equation}
In Minkowski space, it is easy to calculate the electric field generated by the two charges:
\begin{align}\label{charge}
 E_x|_{\rho=\epsilon}&=\frac{Q(x-a)}{r_1^3}-\frac{Q(x-a)}{r_2^3\gamma^2\left(1-v^2 \sin^2\psi\right)^{3/2}}\nn\\
&=\frac{Q(\epsilon \cosh \omega-a)}{r_1^3}-\frac{Q(\epsilon \cosh \omega-a)}{r_2^3\gamma^2\left(1-v^2 \sin^2\psi\right)^{3/2}}\ ,
\end{align}
where,
\begin{align}
r_1^2=& (\epsilon \cosh \omega-a)^2+\left(y-\frac{b}{2}\right)^2+z^2\ ,\\
r_2^2=& (\epsilon \cosh \omega-a)^2+\left(y+\frac{b}{2}-v \epsilon \sinh \omega\right)^2+z^2\ ,\\
\gamma=&\frac{1}{\sqrt{1-v^2}}\ , \qquad \cos \psi= \frac{y+\frac{b}{2}-v \epsilon \sinh \omega}{r_2}\ .
\end{align}
Before we proceed, we can easily check that the total charge on the stretched horizon $Q_{SH}$ is zero for all $\omega$. Let us first imagine a closed surface (in Minkowski space) that encloses both the charges, so that the total charge inside is zero. The Gaussian surface consists of a plane at $x=x_0=\epsilon \cosh \omega$ and a hemisphere at infinity. So, we have,
\begin{equation}
\int dydzE_x|_{x=x_0}=-\int_{\text{infinity}}\vec{E}.\vec{ds} \ .
\end{equation}
It can be easily checked that the quantity in the right hand side vanishes and hence
\begin{equation}
Q_{SH}(\omega)= \frac{1}{4\pi }\int dydzE_x|_{x=\epsilon \cosh \omega}=0\ .
\end{equation}�

\subsubsection{Small velocity limit}
We will now assume that the velocity $v<<1$. As mentioned in the previous section, we want to consider two separate cases: two charges meet inside the Rindler horizon and outside the Rindler horizon. For ease of  computation we will also assume that $b<<1$ so that $b/v$ is of order $\O(1)$. Let us now define the quantity
\begin{equation}
 s= \frac{av}{b}.
\end{equation}
Therefore, $s<1$ means two charges meet inside the Rindler horizon and $s>1$ corresponds to the case when two charges meet outside the Rindler horizon. Therefore, in the small velocity limit, we obtain
\begin{equation}\label{eftotal}
 E_x|_{\rho=\epsilon}=-\frac{3  Q y (a-\epsilon \cosh (\omega)) (a-\epsilon s \sinh (\omega))}{s \left(a^2-2 a \epsilon \cosh (\omega)+\epsilon^2 \cosh ^2(\omega)+y^2+z^2\right)^{5/2}}v+O\left(v^2\right)
\end{equation}
and hence the charge density on the stretched horizon is given by
\begin{equation}\label{chargetotal}
 \sigma(\omega, y , z)=- \frac{3  Q y (a-\epsilon \cosh (\omega)) (a-\epsilon s \sinh (\omega))}{4\pi \epsilon s \left(a^2-2 a \epsilon \cosh (\omega)+\epsilon^2 \cosh ^2(\omega)+y^2+z^2\right)^{5/2}}v+O\left(v^2\right)\ .
\end{equation}
It can be easily checked that the total charge on the stretched horizon is zero. At the leading order, charge density becomes zero everywhere on the stretched horizon at
\begin{align}\label{omegas}
 \cosh (\omega_1)=\frac{a}{\epsilon}\  \qquad \text{and} \qquad \sinh (\omega_2)=\frac{a}{\epsilon s}\ .
\end{align}
It is important to note that at $\omega=\omega_1$,  the charge density is zero everywhere except $y=z=0$, where it becomes infinitely large; we will investigate this more carefully in the next section. We will also show numerically that the small velocity result $\sigma(\omega_2,y,z)=0$, in general, is not true.

Let us now compute the dipole moment of the charge density on the stretched horizon, defined as
\begin{equation}
\vec{p}(\omega)=\int dy dz (y\hat{y}+z\hat{z}) \sigma(\omega, y , z)\ .
\end{equation}
When, $\omega\neq\omega_1$, in the small velocity limit, we obtain
\begin{equation}
\vec{p}(\omega)=\frac{Q v \pi (s \epsilon \sinh (\omega)-a)}{s \epsilon(a-\epsilon \cosh (\omega))}\hat{y}+\O(v^2)
\end{equation}
and hence at the leading order $\vec{p}(\omega_2)=0$. We will show next that this result holds even when the velocity is not small.
\subsubsection{Some general results: dipole moments at $\omega=\omega_1$ and $\omega_2$}
Now let us explore the behavior of the charge distribution on the stretched horizon at $\omega=\omega_1$ and $\omega_2$, when the velocity $v$ is not so small. In the limit $\omega\rightarrow \omega_1$ (but $\omega<\omega_1$), from equation (\ref{charge}), we obtain
\be
E_x|_{\rho=\epsilon}=-2\pi Q \ \delta\left(y-\frac{b}{2}\right)\delta\left(z\right)+2\pi Q \ \delta\left(y+\frac{b}{2}-v \epsilon \sinh \omega_1\right)\delta\left(z\right)\ .
\ee
Therefore, the charge density is not exactly zero. The dipole moment is also nonzero and it is given by,
\be
\vec{p}(\omega_1^{-})=\left(-\frac{Q b}{2 \epsilon}+\frac{v Q \sinh\omega_1}{2}\right)\hat{y}\approx -\frac{Q \left(b-va\right)}{2 \epsilon}\hat{y}\ .
\ee
It is important to note that in the limit $\omega\rightarrow \omega_1$ (but $\omega>\omega_1$), the dipole moment $\vec{p}(\omega_1^{+})=-\vec{p}(\omega_1^{-})$ and hence it is discontinuous at $\omega=\omega_1$.

Now at $\omega=\omega_2$, one can check that in equation (\ref{charge})
\begin{align}
&r_1=r_2=\sqrt{(\epsilon \cosh \omega_2-a)^2+\left(y-\frac{b}{2}\right)^2+z^2}\ ,\\
&\cos \psi= \frac{y-\frac{b}{2}}{r_1}\
\end{align}
and in general $\sigma(\omega_2, y , z)\neq 0$. However, one can show that
\begin{align}
\vec{p}(\omega_2)=&\hat{y}\int dy dz y\sigma(\omega_2, y , z)\nn \\
=& \hat{y}\int d\left(y-\frac{b}{2}\right) dz \left(y-\frac{b}{2}\right) \sigma(\omega_2, y , z)+\hat{y}\frac{b}{2} Q_{SH}\ .
\end{align}
From equation (\ref{charge}), it is clear that $\sigma(\omega_2, y , z)$ is an even function of $y-b/2$ and hence the first term on the right hand side vanishes. The last term on the right hand side is zero because $Q_{SH}=0$. Therefore,
\begin{align}
\vec{p}(\omega_2)=0\ .
\end{align}
Before we proceed few comments are in order.  $\omega=\omega_1$ corresponds to the fact that two charges cross the horizon. Whereas, $\omega_2$ is related to the fact that the charges will meet at some point. When two charges meet inside the Rindler wedge (i.e. $s>1$), one can check that $\omega_2<\omega_1$. When two charges meet outside the Rindler wedge ($s<1$), we will have $\omega_2>\omega_1$. The Rindler observer will never see the charges meet. However, using the available information, it is possible for the Rindler observer to find out the Minkowski time at which two charges will meet
\begin{equation}
t=a\left(\frac{\sinh \omega_2}{\cosh \omega_1}\right)\sim a \ e^{(\omega_2-\omega_1)}\ .
\end{equation}

\subsubsection{Arbitrary velocity}
The electric field (\ref{eftotal}) is a linear superposition of a Coulombic profile (for the static charge) and a  boosted version of the same (for the moving charge). Each of these configurations is invariant under $z\leftrightarrow -z$ so as a result, the component of the dipole moment $p_z$ vanishes identically for all $t$ (and all $\omega$). On the other hand, the electric field and induced charge density for the charge $Q$ are invariant under $y-b/2\leftrightarrow -y+b/2$, but for the charge $-Q$ the same is true if $y+b/2-vt\leftrightarrow -y-b/2+vt$.

The superposition of these two profiles does not respect any kind of parity in $y$, but for the special time $t=b/v$ the whole configuration is invariant under $y-b/2\leftrightarrow -y+b/2$. Then, we conclude that the dipole moment $\vec{p}$ measured with respect to $(b/2,0)$ vanishes for this particular time. Moreover, because $Q_{SH}=0$ then the same holds true for any other choice of the coordinate origin and, in particular, for $(0,0)$. In terms of the Rindler time, this happens when
\be
\omega_2=\sinh^{-1}\left(\frac{b}{\epsilon v}\right)\simeq\log\left(\frac{2b}{\epsilon v}\right)+\mathcal{O}(\epsilon^2)\,.
\ee
Thus, we have verified that our formula for the time for which the dipole moment vanishes (also obtained in (\ref{omegas})) holds for arbitrary velocity. However, the charge density is not zero for arbitrarily high velocities. This is because the boosted Coulombic profile corresponding to the charge $-Q$ only cancels the static profile of the charge $Q$ in the $v\ll1$ limit.

\begin{figure}[h]
\begin{center}
\unitlength = 1mm

\subfigure[ ]{
\includegraphics[width=45mm]{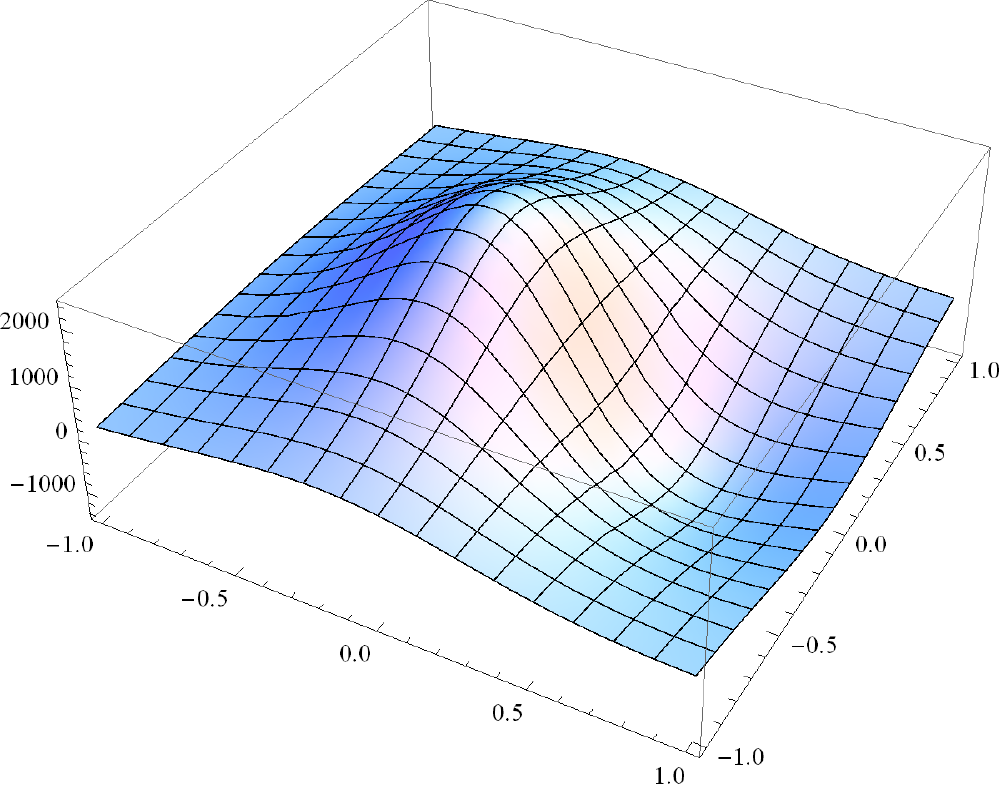}
}
\subfigure[ ]{
\includegraphics[width=45mm]{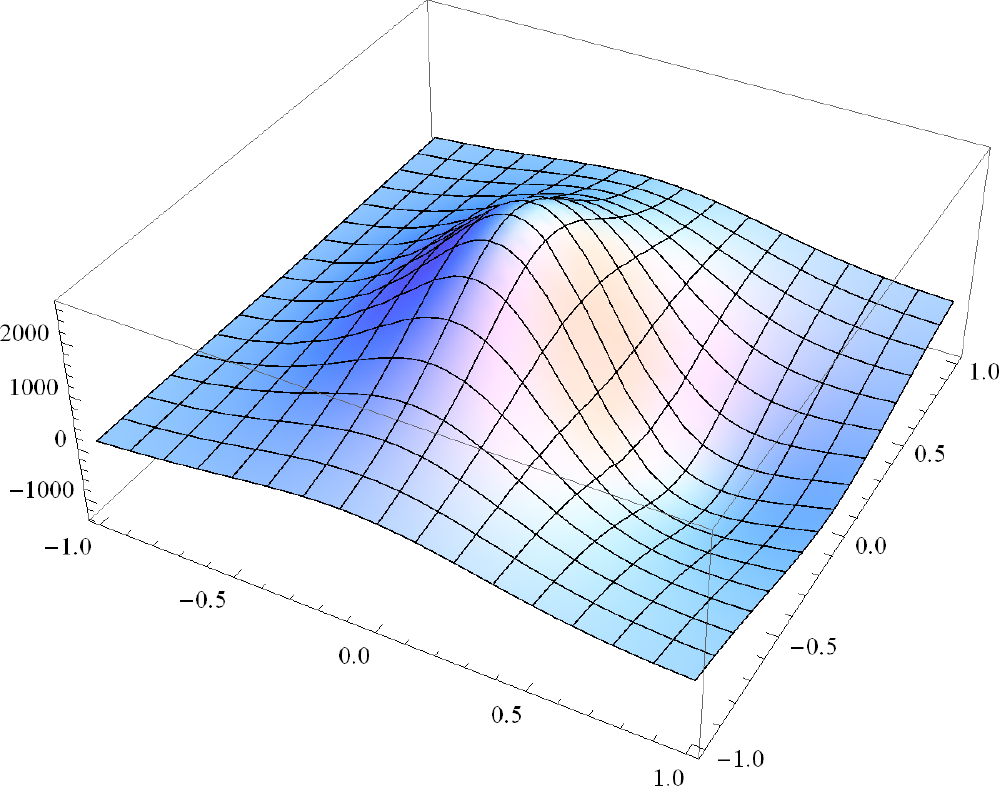}
 }
\subfigure[ ]{
\includegraphics[width=45mm]{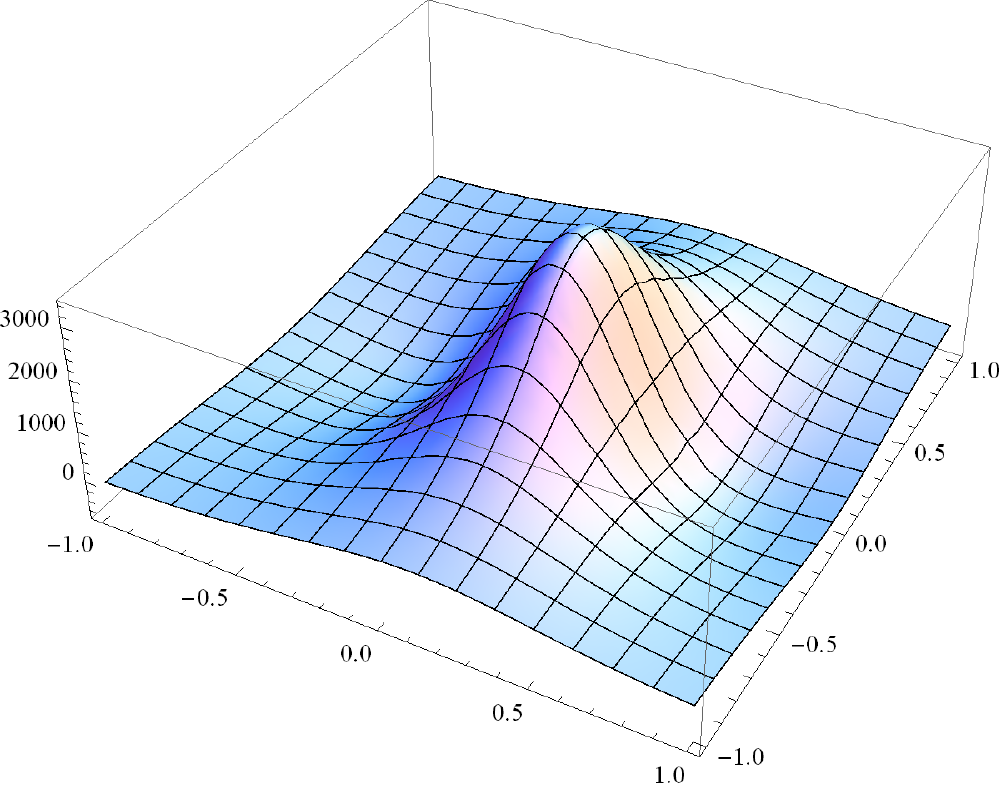}
 }
\caption{\small Time evolution of the charge density $\sigma(y,z)$ for a situation in which the two charges meet before crossing the horizon. For the plots we chose the following parameters: $Q=1$, $\epsilon=10^{-5}$, $a=1$, $b=0.2$, $v=0.5$ and (a) $\omega=0.9\times\omega_2$ (b) $\omega=0.95\times\omega_2$ (c) $\omega=\omega_2$.} \label{fig2}
\end{center}
\end{figure}

\begin{figure}[h]
\begin{center}
\unitlength = 1mm

\subfigure[ ]{
\includegraphics[width=45mm]{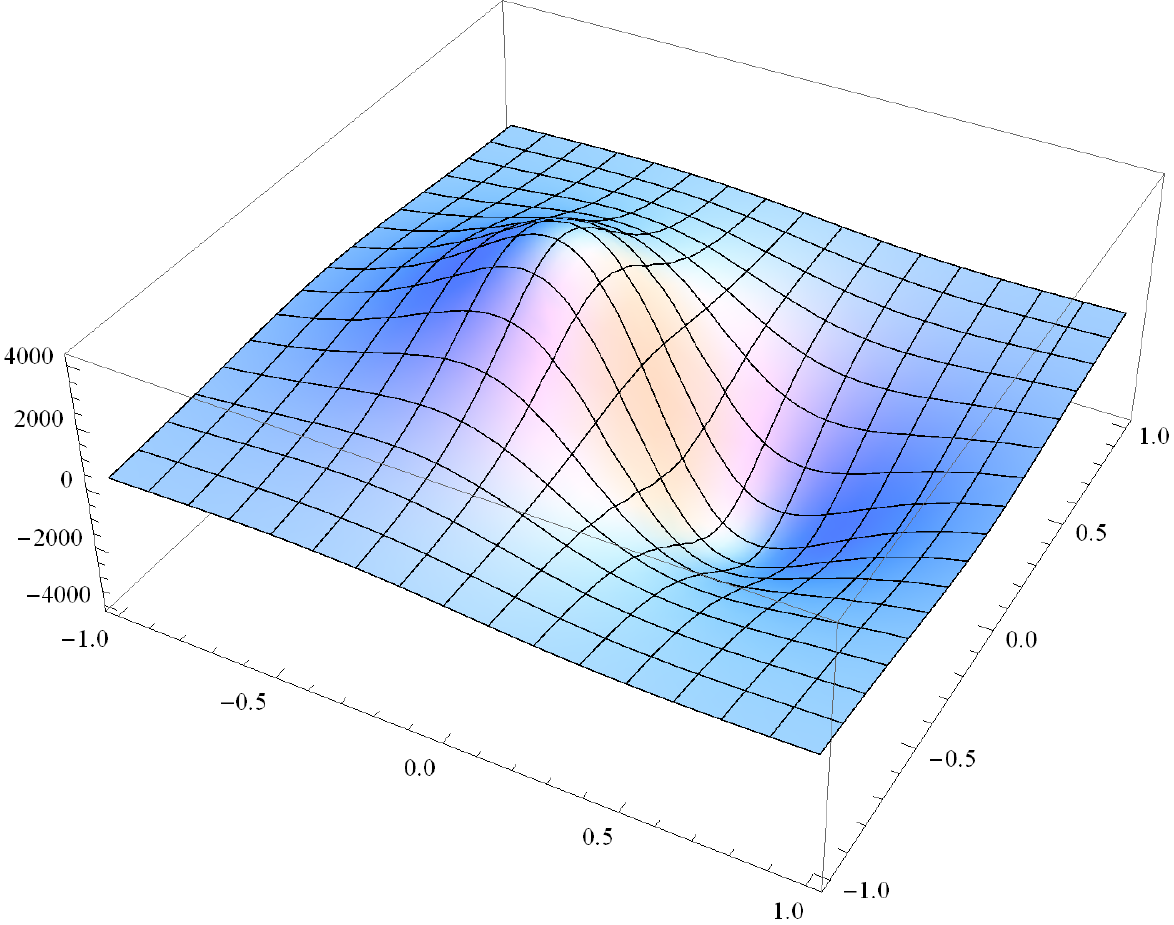}
}
\subfigure[ ]{
\includegraphics[width=45mm]{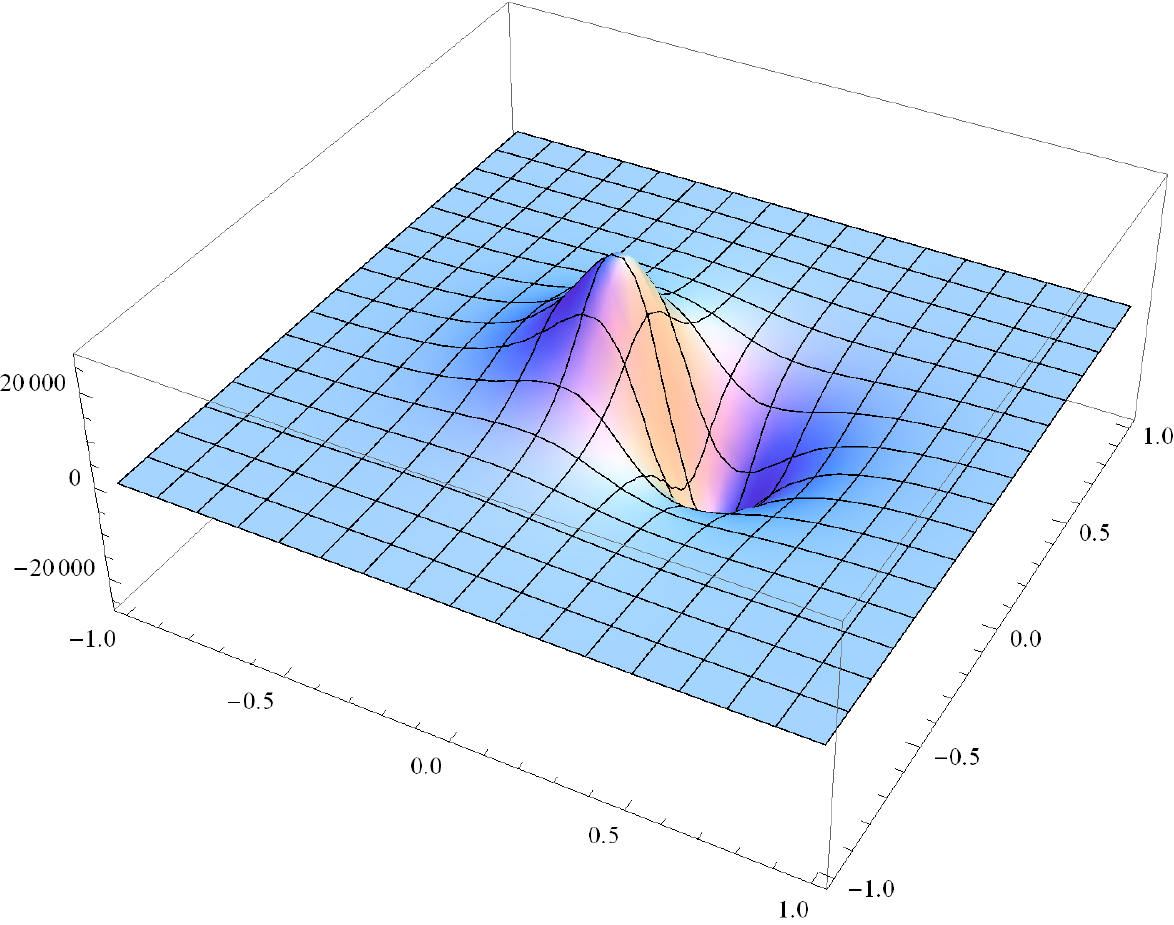}
 }
\subfigure[ ]{
\includegraphics[width=45mm]{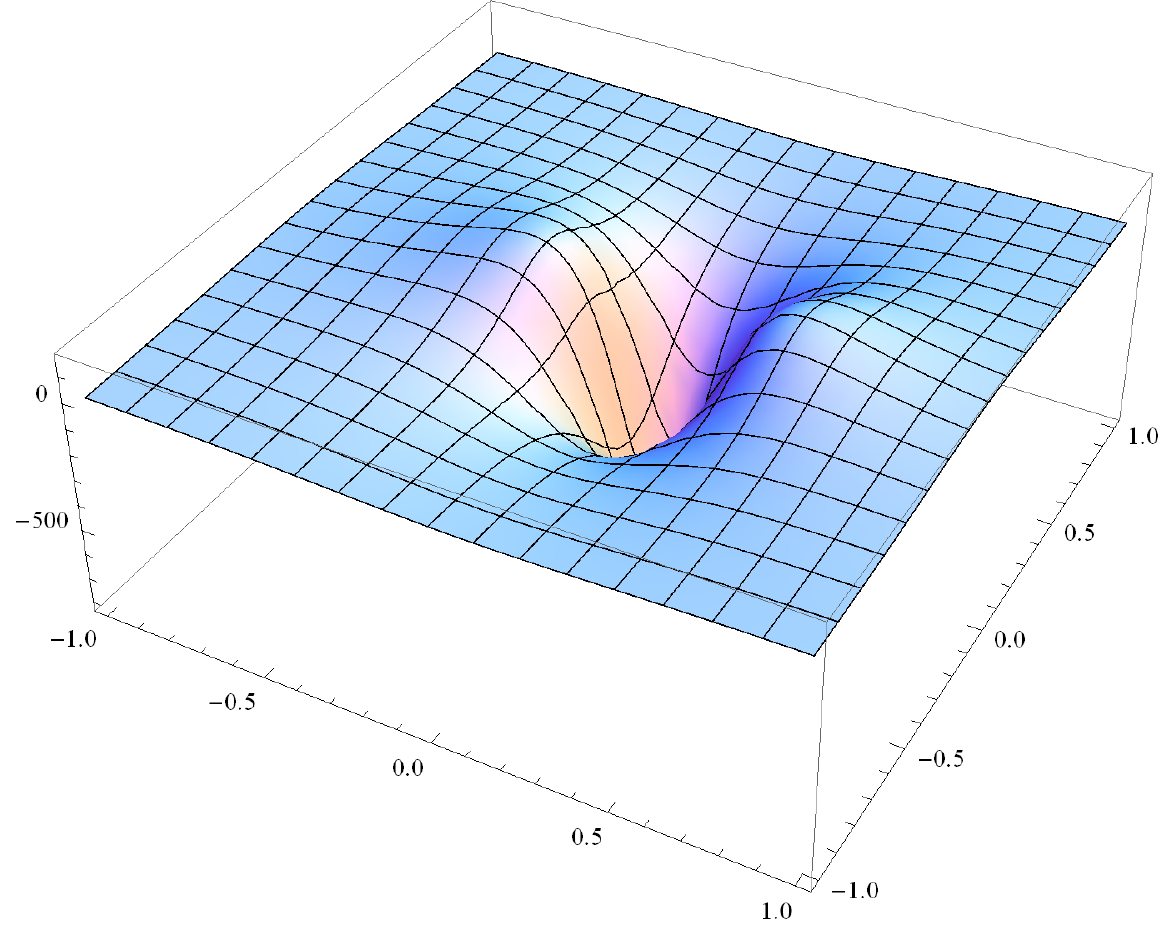}
 }
\caption{\small Time evolution of the charge density $\sigma(y,z)$ for a situation in which the two charges meet after crossing the horizon. For the plots we chose the following parameters: $Q=1$, $\epsilon=10^{-5}$, $a=1$, $b=0.2$, $v=0.15$ and (a) $\omega=0.9\times\omega_2$ (b) $\omega=0.95\times\omega_2$ (c) $\omega=\omega_2$. For this set of parameters $\omega_1\approx0.977\times\omega_2$ and therefore the charge density in (c) appears inverted with respect to the first two plots.} \label{fig3}
\end{center}
\end{figure}

\begin{figure}[h]
\begin{center}
\unitlength = 1mm

\subfigure[ ]{
\includegraphics[width=60mm]{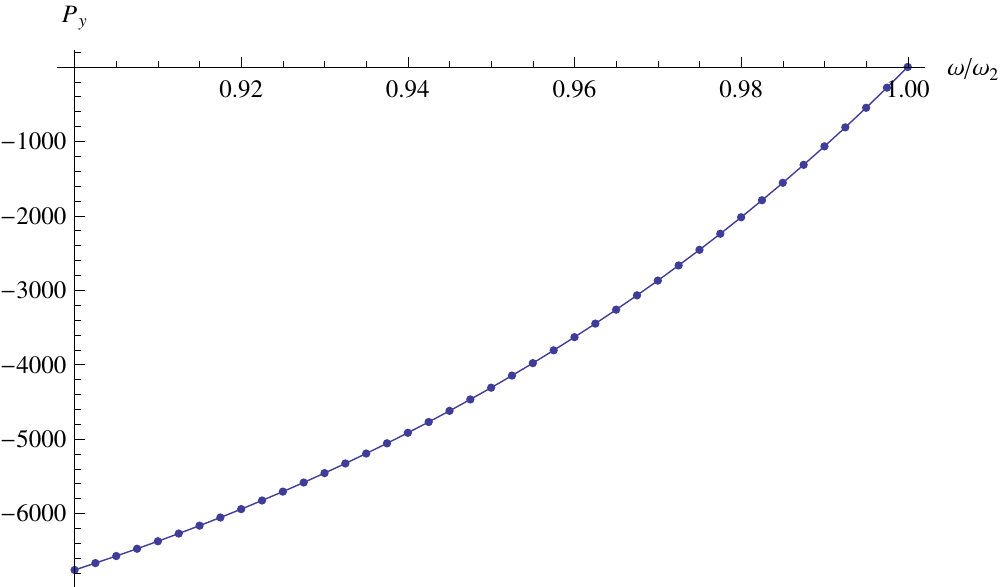}
}
\subfigure[ ]{
\includegraphics[width=60mm]{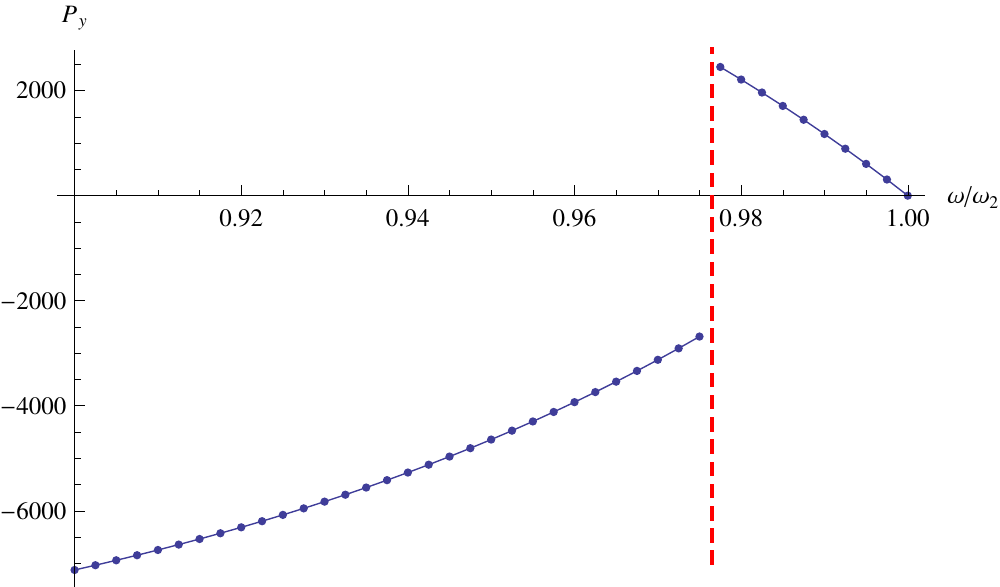}
 }
\caption{\small Time evolution of the dipole moment $p_y$ as a function of $\omega/\omega_2$ for (a) the configuration shown in Figure \ref{fig2}, where two charges meet inside the Rindler wedge (b) the configuration shown in Figure \ref{fig3}, where two charges meet after they cross the horizon. For this last configuration, and for the set of parameters we have chosen, it is found that $\omega_1\approx0.977\times\omega_2$ (depicted in red).} \label{fig4}
\end{center}
\end{figure}

In Figures \ref{fig2} and \ref{fig3} we plot three consecutive snapshots of the induced charge density $\sigma(y,z)$ for $\omega=0.9\times\omega_2$, $0.95\times\omega_2$ and $\omega_2$. Figure \ref{fig2} represents a case in which the two charges meet inside the right Rindler wedge, whereas in Figure \ref{fig3} the two charges meet after they cross the horizon. In both cases the membrane dynamics correctly accounts for the fact that the total dipole $\vec{p}$ vanishes at $\omega=\omega_2$, in which time the charge density respects the parity symmetries discussed above. In figure \ref{fig4} we integrate numerically the component of the dipole moment $p_y$ as a function of time for these two configurations. As expected the two of them go to zero as $\omega\to\omega_2$. However, only in the second case we can observe the jump in the dipole moment expected for $\omega=\omega_1$, \emph{i.e.} when the two charges cross the horizon.

\subsection{Smearing of charges for a Schwarzschild black hole}
The metric of a Schwarzschild black hole in Kruskal-Szekeres coordinates is given by,
\begin{equation}
ds^2=\frac{32G^3m^3}{r}e^{-\frac{r}{2Gm}}\left(-dV^2+dU^2\right)+r^2 \left(d\theta^2 + \sin^2\theta d\phi^2\right)\ ,
\end{equation}
where $r$ is the radial Schwarzschild coordinate
\begin{equation}
V^2-U^2=\left(1-\frac{r}{2Gm}\right)e^{\frac{r}{2Gm}}\ .
\end{equation}
The event horizon is located at $V=\pm U$ and the curvature singularity is at $V^2-U^2=1$.
\begin{figure}[!htbp]
\begin{center}
  \includegraphics[width=9.5cm]{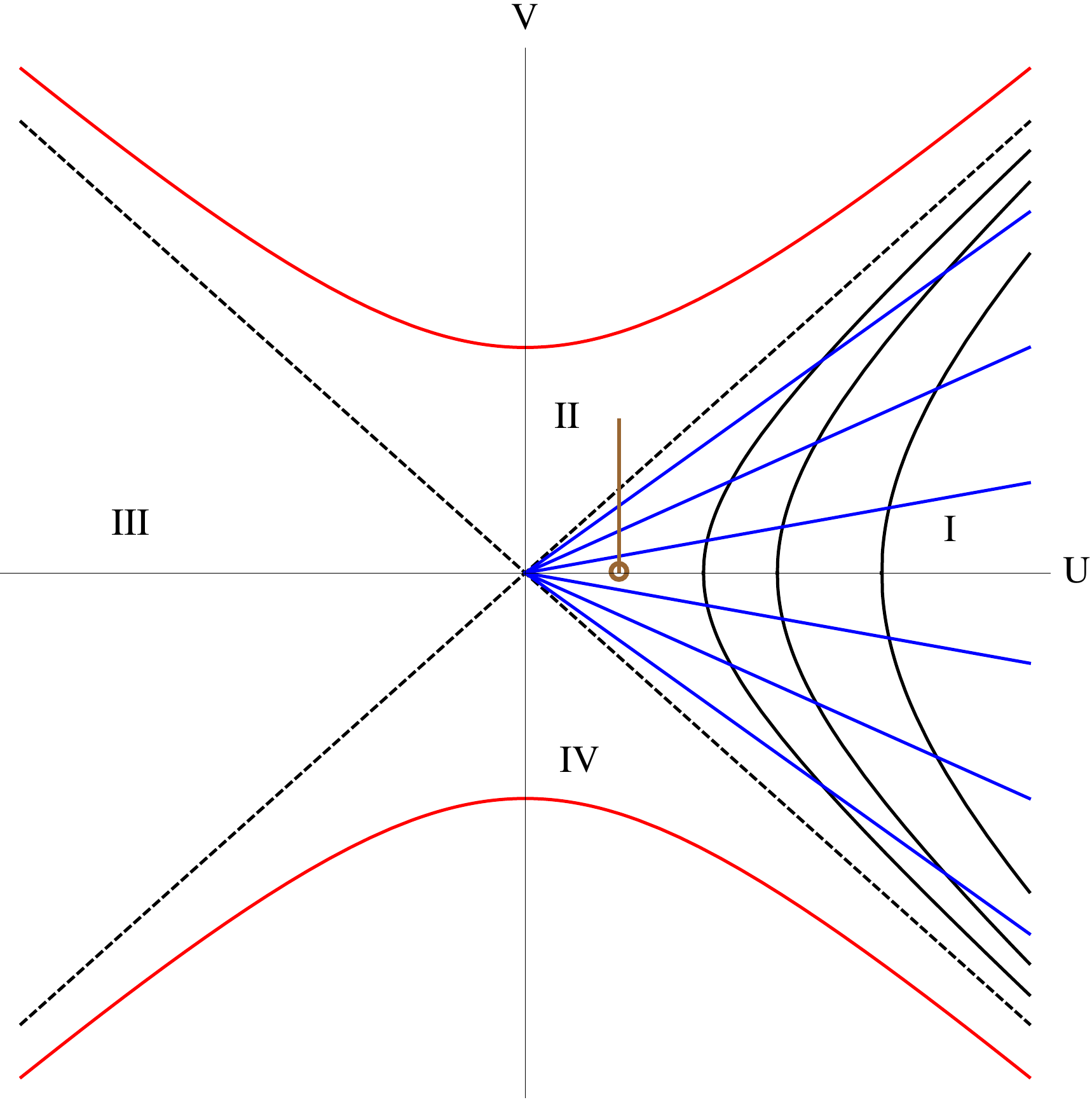}
  \end{center}
\caption{Maximal analytic extension of Schwarzschild solution in Kruskal-Szekeres coordinates. Dashed black lines that divide the diagram in four regions are the event horizons. Region I is the Schwarzschild patch, where constant $r$-lines are shown in black and constant $t$-lines are shown in Blue. The solid red lines represent the singularity. The world line of a free falling, time dependent dipole (perpendicular to the plane) from $V=0$ to the time when two charges meet, is shown in brown.}
\end{figure}
In Schwarzschild coordinates, the metric is given by,
\begin{equation}
ds^2=-\left(1-\frac{r}{2Gm}\right)dt^2+\frac{dr^2}{\left(1-\frac{r}{2Gm}\right)}+r^2 \left(d\theta^2 + \sin^2\theta d\phi^2\right)\ .
\end{equation}
For $r>2Gm$, coordinates $\{U,V\}$ and $\{r,T\}$ are related in the following way
\begin{align}
V=& \left(\frac{r}{2Gm}-1\right)^{1/2}e^{\frac{r}{4Gm}}\sinh \left(\frac{t}{4Gm}\right)\ , \\
U=& \left(\frac{r}{2Gm}-1\right)^{1/2}e^{\frac{r}{4Gm}}\cosh \left(\frac{t}{4Gm}\right)\ .
\end{align}
Following our procerdure in the Rindler case, we will replace the mathematical horizon by the stretched horizon at $r=2Gm+\delta$, where $\delta$ is of the order of Planck length.

Let us now consider a large black hole and restrict the charges to be near the horizon, i.e. $|r/2Gm-1|<<1$, in a small angular region arbitrarily centered at $\theta=0$. In that case, we can replace the angular part of both Kruskal-Szekeres and Schwarzschild coordinates by Cartesian coordinates
\begin{equation}
r^2 \left(d\theta^2 + \sin^2\theta d\phi^2\right)\approx dy^2+dz^2\ ,
\end{equation}
where,
\begin{align}
y=2mG \theta \cos\phi\ ,\qquad z=2mG \theta \sin\phi\ .
\end{align}
We will now define two sets of coordinates to describe the near horizon region:
\begin{align}
&\rho=4Gm\sqrt{1-\frac{2Gm}{r}}, \qquad \omega=\frac{t}{4Gm}\ .\\
&X= 4Gme U, \qquad T= 4Gme V\ .
\end{align}
Where $e$ is the Euler's number. Therefore, in terms of $\{T,X,y,z\}$, the Kruskal-Szekeres metric becomes (near the horizon)
\begin{equation}
ds^2\approx-dT^2+dX^2+dy^2+dz^2\ .
\end{equation}
Similarly, the Schwarzschild metric leads to
\begin{equation}
ds^2\approx-\rho^2d\omega^2+d\rho^2+dy^2+dz^2\ .
\end{equation}
Therefore, we can use the results of the previous section to analyze smearing of charges on the Schwarzschild horizon. Again we have two charges $Q$ and $-Q$ and initial conditions are given at $V=0$:
\begin{align}
Q:& \qquad U=U_0,\ \theta=\frac{\theta_0}{2}, \ \phi=0. \\
-Q:& \qquad U=U_0,\ \theta=-\frac{\theta_0}{2}, \ \phi=0, \frac{d\theta}{dV}=\alpha.
\end{align}
In the near horizon approximation, the trajectory of the charge $-Q$ is given by: $\theta=-\theta_0/2+\alpha V$ and hence two charges will meet at the Kruskal-Szekeres time $V=\theta_0/\alpha$. The Schwarzschild observer will never see two charges meet if $V=\theta_0/\alpha>U_0$. However the discussion of the last section indicates that the Schwarzschild observer has the information that the two charges will meet inside the black hole horizon. In the near horizon approximation, we can map this problem to the problem of the last section, with
\begin{align}
a=4GmeU_0, \qquad b=&2Gm\theta_0, \qquad v=\frac{\alpha}{2e}, \qquad \epsilon=2\sqrt{2Gm\delta}\ ,\\
&s=\frac{av}{b}=\frac{U_0\alpha}{\theta_0}\ .\nn
\end{align}
Therefore, the Schwarzschild observer can see the two charges for $t<t_1$. At Schwarzschild time $t=t_1$, the Schwarzschild observer will see that the charge density on the stretched horizon is localized at two points , where
\be
\cosh\left(\frac{t_1}{4Gm}\right)=\frac{e U_0 \sqrt{2Gm}}{\sqrt{\delta}}\ .
\ee
At a later time $t=t_2$, the dipole moment of the charge distribution on the stretched horizon becomes zero
\be
\sinh\left(\frac{t_2}{4Gm}\right)=\frac{e U_0\sqrt{2Gm}}{s\sqrt{\delta}}\ .
\ee
Using these information, it is possible for the Schwarzschild observer to find out the Kruskal-Szekeres time $V$ at which two charges will meet
\be
V=U_0\left(\frac{\sinh(t_2/4Gm)}{\cosh(t_1/4Gm)}\right)\sim U_0 e^{\frac{t_2-t_1}{4Gm}}\ .
\ee

It is important to point out that there is a crucial difference between the discussions of the last two sections. In a Schwarzschild black hole, all freely falling objects will hit the singularity at $r=0$ in finite  Kruskal-Szekeres time. One can perhaps argue that there might be some signature on the stretched horizon when a free falling charge hits the singularity. Let us assume that two freely falling charges, $+Q$ and $-Q$, separated by an arbitrary angle $\theta$ hit the singularity at the same time. Discussion of this section indicates that the Schwarzschild observer, as a consequence of that, should find the dipole moment of the charge distribution on the stretched horizon to be zero after some Schwarzschild time. Therefore, it is reasonable to expect that when a single charge hits the singularity, the spherical symmetry will be restored and the total charge will be uniformly distributed over the stretched horizon.

More generally, we can argue from the proof of the No Hair Theorem\cite{nohair} that all perturbations of the fields on the horizon decay on a time scale of the light crossing time of the horizon.  This gives the smallest imaginary part for the frequencies of quasi-normal modes.  In the scrambling time, $2 G m {\rm ln}\ (\frac{2m}{M_P}) $ an order one perturbation will decay to size $ \sim \frac{M_P}{m} $, and all trace of it will be lost \cite{fastscramble} \footnote{In fact, several authors have argued that non-locality is indeed an essential property of fast scramblers \cite{nonlocal}, a feature that is not present in QUEFT. This is further supported by the fact that non-local interactions increase the level of entanglement among the different degrees of freedom of the theory \cite{entanglement}.}. This should be the time scale for the classical fields on the horizon to become spherically symmetric. A more detailed calculation in the full Kruskal-Szekeres background would undoubtedly confirm this conclusion based on general theorems, but we will not attempt it in this paper.

\vskip.3in
\begin{center}
{\bf Acknowledgments }
\end{center}
T.B. would like to acknowledge conversations with J. Polchinski, L.Susskind, J. Maldacena and D.Harlow about firewalls.  The work of T.B. was supported in part by the Department of Energy.   The work of W.F. was supported in part by the TCC and by the NSF under Grant PHY-0969020

\end{document}